\documentclass{article}

\usepackage{microtype}
\usepackage{graphicx}
\usepackage{subcaption}
\usepackage{booktabs} 

\usepackage{hyperref}
\usepackage{amsfonts} 
\usepackage{amsmath}
\usepackage{amsthm}
\usepackage{amssymb}
\usepackage{bm}
\usepackage{tabulary}
\usepackage{tablefootnote}
\usepackage{multirow}
\newcommand{\printfnsymbol}[1]{%
  \textsuperscript{\@fnsymbol{#1}}%
}

\newtheorem{theorem}{Theorem}
\newtheorem{mydef}{Definition}
\newtheorem{lemma}{Lemma}
\usepackage[accepted]{icml2021}

\icmltitlerunning{Deoscillated Adaptive Graph Collaborative Filtering}

\begin{document}

\twocolumn[
\icmltitle{Deoscillated Adaptive Graph Collaborative Filtering}



\icmlsetsymbol{equal}{*}

\begin{icmlauthorlist}
\icmlauthor{Zhiwei Liu}{go,equal}
\icmlauthor{Lin Meng}{to,equal}
\icmlauthor{Fei Jiang}{the}
\icmlauthor{Jiawei Zhang}{to}
\icmlauthor{Philip S. Yu}{go}
\end{icmlauthorlist}

\icmlaffiliation{go}{Department of Computer Science, University of Illinois at Chicago, USA}
\icmlaffiliation{to}{Department of Computer Science, Florida State University, USA}
\icmlaffiliation{the}{Committee on Computational and Applied Mathematics, University of Chicago, USA}

\icmlcorrespondingauthor{Zhiwei Liu}{zliu213@uic.edu}
\icmlcorrespondingauthor{Lin Meng}{lin@ifmlab.org}
\icmlcorrespondingauthor{Fei Jiang}{feijiang@uchicago.edu}
\icmlcorrespondingauthor{Jiawei Zhang}{jiawei@ifmlab.org}
\icmlcorrespondingauthor{Philip S. Yu}{psyu@uic.edu}

\icmlkeywords{Machine Learning, ICML}

\vskip 0.3in
]



\printAffiliationsAndNotice{\icmlEqualContribution} 

\begin{abstract}
Collaborative Filtering~(CF) signals are crucial for a Recommender System~(RS) model to learn user and item embeddings. High-order information can alleviate the cold-start issue of CF-based methods, which is modeled through propagating the information over the user-item bipartite graph. Recent Graph Neural Networks~(GNNs) propose to stack multiple aggregation layers to propagate high-order signals. 
However, there are three challenges, the oscillation problem, varying locality of bipartite graph, and the fixed propagation pattern, which spoil the ability of the multi-layer structure to propagate information.
In this paper, we theoretically prove the existence and boundary of the oscillation problem, and empirically study the varying locality and layer-fixed propagation problems. We propose a new RS model, named as \textbf{D}eoscillated adaptive \textbf{G}raph \textbf{C}ollaborative \textbf{F}iltering~(DGCF), which is constituted by stacking multiple CHP layers and LA layers.
We conduct extensive experiments on real-world datasets to verify the effectiveness of DGCF. Detailed analyses indicate that DGCF solves oscillation problems, adaptively learns local factors, and has layer-wise propagation patterns. 
Our code is available online\footnote{\url{https://github.com/JimLiu96/DeosciRec}}.

\end{abstract}

\section{Introduction}

A Recommender System~(RS)~\cite{rendle2009bpr,rendle2010factorization,he2017neural,wang19neural,zheng17joint,pinsage2018ying} can predict the interests of users towards items, where a typical method is to model the Collaborative Filtering~(CF) signals~\cite{wang19neural}. Assuming similar users share relevant items, CF signals help to learn the user and item embeddings given user-item interactions. One major issue regarding CF methods is the cold-start problem~\cite{zheng2018spectral}, where embeddings of users or items with few interactions are hard to learn. To remedy this issue, we can model \textit{high-order signals}~\cite{wang19neural} that propagate information from multiple hops away over the user-item bipartite graph. Recently, owing to the development of Graph Neural Network~(GNN)~\cite{kipf17semi,graphsage17hamilton,wu19simplifying}, we can easily propagate high-order signals by stacking multiple \textit{aggregation layers}. Therefore, users (items) can aggregate the first-order signal from direct neighbors at the first layer and high-order signals from indirect neighbors at deep layers. However, via experimental analyses on real-world datasets, we observe three common performance problems of existing GNNs on RS, which are named as the \textbf{oscillation} problem, \textbf{varying locality} problem, and \textbf{fixed propagation pattern} problem in this paper, respectively.

\begin{figure}
    \centering
    \includegraphics[width=0.99\linewidth]{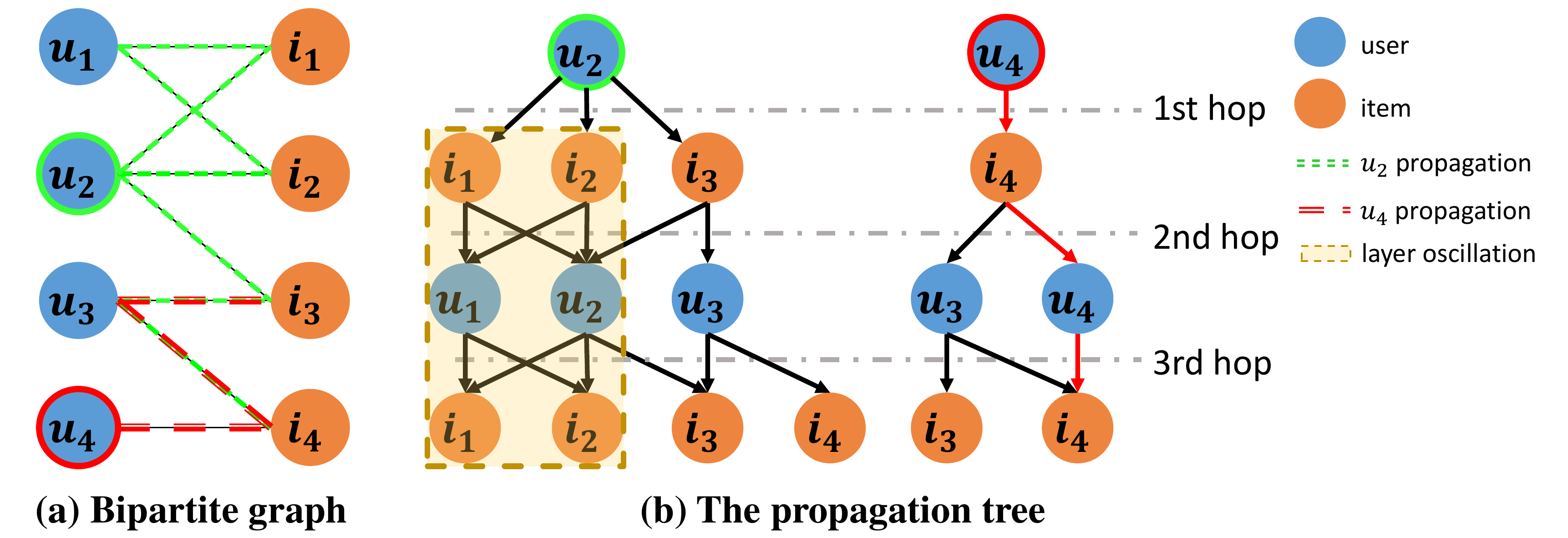}
    \vspace{-2mm}
    \caption{The left part is an example of bipartite graph. The green and red color labels the edges covered by 3-hop propagation of $u_2$ and $u_4$, respectively. The right part presents the corresponding 3-hop propagation tree of $u_2$ and $u_4$. }
    \label{fig:example_osci}
\end{figure}

Firstly, stacking multiple layers leads to the oscillation problem, which in turn hinders the information to propagate. Before formally defined, intuitively, oscillation problem occurs if there is an information gap between the propagation of successive hops. It results from the bipartite structure. As illustrated in the left part of Figure~\ref{fig:example_osci}, the direct neighbors of users are all items, while the 2-hop neighbors are all users. This implies that by aggregating the direct neighbors, users only receive the information from items, and vice versa.  It turns out that the information oscillates between users and items due to the bipartite graph structure. One quick solution is adding self-loops~\cite{kipf17semi,wang19neural} to break the bipartite structure. However, we theoretically prove that it cannot alleviate the oscillation problem. Additionally, adding self-loops is equivalent to not propagating the information, which even exacerbates the cold-start issue. 


Besides the oscillation problem, the varying locality problem of bipartite graph also limits the propagating ability of the multi-layer structure~\cite{xu18jk}. It refers to the density of local structures over the bipartite graph. For example, on the left-hand side of Figure~\ref{fig:example_osci}, $u_4$ has a lower local density compared with $u_2$ as the former only directly connects to $i_4$ while the latter connects to $i_1$, $i_2$ and $i_3$. The $3$-hop propagation starting from $u_4$ (red edges in Figure~\ref{fig:example_osci}) only covers $3$ edges in the bipartite graph. In contrast, the $3$-hop propagation of $u_2$ (green edges in Figure~\ref{fig:example_osci}) covers almost every edge in the bipartite graph. Specifically, we present the corresponding propagation tree of $u_2$ and $u_4$ on the right-hand side of Figure~\ref{fig:example_osci}. There are fewer possible propagation paths for $u_4$ compared with $u_2$. 


Last but not least, the fixed propagation pattern at different layers of existing multi-layer RS models~\cite{he2020lightgcn,wang19neural,berg2017gcmc} induces redundant information propagation between layers. We present an example of the redundancy as the yellow-shaded block on the right-hand side of Figure~\ref{fig:example_osci}. The first hop propagation distributes the information of $u_2$  to $i_1$ and $i_2$. Then, the information from $i_1$ and $i_2$ is propagated to $u_1$ and $u_2$ at the second hop propagation, which is reversely propagated back at the third hop propagation. This is due to the propagation pattern is fixed at different layers. One possible solution is to sample layer-dependent neighbors~\cite{zou2019layer}. However, the current sampling strategy requires additional computation and is not differentiable, which spoils the training procedure. 


In order to tackle the aforementioned problems, we design a new GNN framework specifically for bipartite graphs, named deep \textbf{D}eoscillated Adaptive \textbf{G}raph \textbf{C}ollaborative \textbf{F}iltering~(DGCF). We propose a Cross-Hop Propagation~(CHP) layer that  remedies the bipartite propagation structure to resolve the oscillation problem. Compared with existing methods that aggregate only direct neighbors, the CHP layer also propagates the information from cross-hop neighbors. As a result, users (items) can also receive the information from cross-hop users (items). Moreover, we design a Locality-Adaptive~(LA) layer which controls the amount of information to propagate. It learns an influencing factor for each node, which is adaptive to the varying locality of the graph. Multiple CHP layers and LA layers constitute the DGCF model. It propagates high-order CF signals to train user embeddings and item embeddings, which is adaptive to the locality of nodes. The weights of LA layers and CHP layers are different from layer to layer, thus being a layer-wise propagation pattern. The contributions are:
\begin{itemize}
\item \textbf{Oscillation problem:} We formally define the graph oscillation problem and prove its existence and boundary in bipartite graphs. We also provide empirical evidence that oscillation occurs with a few layers. 
    
\item \textbf{Innovative layers:} We design two innovative layers, i.e., CHP layer and LA layer. CHP layer is designed to overcome oscillation problem, while LA layer is for varying locality problem. Stacking multiple LA layers resolves the fixed aggregation pattern problem.
    
\item \textbf{Extensive experiments:} In addition to the overall comparison experiment, we conduct three types of detailed analyses regarding the three aforementioned problems. We study the oscillation problem when stacking too many layers. Additionally, we discuss the varying locality problem and how our model resolves it. Moreover, we observe the variations of the propagation pattern. 
\end{itemize}

\section{Preliminary and Related Work}
Nowadays, Graph Neural Networks (GNNs)~\cite{kipf17semi,wu19simplifying,liu2020alleviating} are widely used in recommender system~(RS)~\cite{he2020lightgcn,pinsage2018ying,wang19neural}. In this section, we review the concepts of GNNs and their association with RS. 

\subsection{Graph Neural Network}
Given a simple graph $\mathcal{G} = (\mathcal{V}, \mathcal{E})$, where $\mathcal{V}$ and $\mathcal{E}$ denote the nodes and edges, respectively. The input feature of node $v$ can be denoted as $\mathbf{x}_v\in \mathbb{R}^{d}$ and the hidden feature learned at $l$-th layer is denoted by $\mathbf{h}^{(l)}_v\in\mathbb{R}^{h}$, where $d$ is dimension of the input feature, and $h$ is the dimension of the hidden features. We define $\mathbf{h}_v^{(0)}=\mathbf{x}_v$. Generally, GNN models have aggregation layer that aggregates features of selected neighboring nodes
. The information is propagated from the selected nodes. Note that `aggregation' describes the information diffusion from the view of target nodes, while `propagation' is from the view of source nodes. Hence, we use them interchangeably in the remaining parts. Formally, the general propagation layer can be denoted by:
\begin{equation}
    \mathbf{h}_{v}^{(l)} = \sigma\left(\sum_{u\in \mathcal{N}_v\cup\{v\}}\alpha_{vu}\mathbf{h}_u^{(l-1)}\mathbf{W}^{(l-1)}\right),
\end{equation}
where the set $\mathcal{N}_v=\{u\in\mathcal{V}|(u,v)\in\mathcal{E}\}$ is the selected neighboring node set of node $v$, and $\mathbf{W}^{(l)}$ denotes a linear transformation, and $\sigma$ denotes a non-linear activation function (e.g., ReLU). $\alpha_{uv}$ denotes the coefficient between node $v$ and neighboring node $u$. Particularly, GCN directly aggregates all the neighboring nodes:
\begin{equation}\label{eq:GCN}
    \mathbf{H}^{(l)} = \sigma\left(\Tilde{\mathbf{A}}\mathbf{H}^{(l-1)}\mathbf{W}^{(l-1)}\right),
\end{equation}
where $\Tilde{\mathbf{A}}$ is defined as the Laplacian normalized adjacency matrix $\Tilde{\mathbf{A}}=\mathbf{D}^{-\frac{1}{2}}(\mathbf{A}+\mathbf{I})\mathbf{D}^{-\frac{1}{2}}$, and $\mathbf{H}^{(0)}=\mathbf{X}$. The GCN layer can learn its 1-st hop neighbors, which can be also seen as one step information propagation from the starting node. By stacking $L$ layers, 
node features propagate to the nodes on the $L$-th hop. Note that $\Tilde{\mathbf{A}}$ can also be random walk normalization, i.e., $\Tilde{\mathbf{A}}=\mathbf{D}^{-1}\mathbf{A}$~\cite{Zhang2019GResNetGR}. The nonlinear activation between layers and redundant linear transformations are removed later in SGC~\cite{wu19simplifying}, $\mathbf{H}^{(l)} =\Tilde{\mathbf{A}}^{L}\mathbf{X}\mathbf{W}.$ Instead of stacking $L$ layers, SGC directly calculates the $L$ powers of $\Tilde{\mathbf{A}}$ as multi-hop aggregation, which greatly reduces the number of linear transformation matrix to one rather than $L$. 


\subsection{Recommender System}
In typical Recommender Systems~(RS), user-item interactions can be represented as a user-item bipartite graph. Meanwhile, GCNs shows unprecedented representation power on many areas including RS, where the collaborative filtering signal can be modeled via the high-order information propagation. Recently, NGCF~\cite{wang19neural} is proposed specifically for recommendation, which successfully models the collaborative signal in bipartite graphs. Different from GCN layers, a NGCF propagation layer additionally includes dot product to model the propagation messages, which can be denoted as:
\begin{equation}
    \mathbf{E}^{(l)}=\sigma\left(\Tilde{\mathbf{A}} \mathbf{E}^{(l-1)} \mathbf{W}_{1}^{(l)}+ \hat{\mathbf{A}} \mathbf{E}^{(l-1)} \odot \mathbf{E}^{(l-1)} \mathbf{W}_{2}^{(l)}\right)
\end{equation}
where $\mathbf{W}_{1}^{(l)}, \mathbf{W}_{2}^{(l)}$ are linear mapping matrices of the hidden embeddings. $\mathbf{E}^{(l)}$ denotes the embeddings for both users and items at $l-$th layer. And $\hat{\mathbf{A}}=\mathbf{D}^{-\frac{1}{2}}\mathbf{A}\mathbf{D}^{-\frac{1}{2}}$, which is different from $\Tilde{\mathbf{A}}$ by the self-loop. To reduce the time and space complexity, LightGCN~\cite{he2020lightgcn} is proposed, which shows that a propagation layer without dot product, and redundant linear transformations even improves performance. Formally, a propagation layer of LightGCN is:
\begin{equation}
    \mathbf{E}^{(l)}= \hat{\mathbf{A}}\mathbf{E}^{(l-1)}.
\end{equation}
There are other works related to GCN based RS. GC-MC~\cite{berg2017gcmc} predicts links between users and items by applying the graph convolutional network as the graph encoder.  GraphSage~\cite{graphsage17hamilton} learns the graph embedding by aggregating the information of neighbors, which is extended as a large-scale recommender system, namely PinSage~\cite{pinsage2018ying}. SpectralCF~\cite{zheng2018spectral} designs a spectral convolutional filter to model the CF signals in user-item bipartite graphs. Although existing works~\cite{liu2020basconv} are effective in RS, we find that those models fail to notice the oscillation problem when applying the GNN to bipartite graph.

\subsection{Comparison of GNN and Graph CF}
In this paper, we focus on combining the Graph CF method~\cite{wang19neural,he2020lightgcn} with the GNN research area~\cite{kipf17semi,graphsage17hamilton,wu19simplifying}. The information propagation process is similar in both areas. Therefore, we can use off-the-shelf GNN research findings to design the graph-based CF model. However, most existing GNN models are built upon attribute graphs, while the CF method mainly focuses on learning the latent embeddings purely from bipartite graph structure. This leads to the difference in \textit{trainable parameters} and \textit{final embeddings} of GNN and Graph CF models. The trainable parameters in most GNNs are the convolutional weights. For example, the GCN model only has $\mathbf{W}\in\mathbb{R}^{d\times d}$ being the trainable weight. The initial embeddings $\mathbf{H}^{(0)}$ is the node features. However, existing CF models require trainable initial embeddings of all nodes, i.e., $\mathbf{E}^{(0)}\in\mathbb{R}^{N\times d}$ is the trainable parameters~\cite{zheng2018spectral,he2020lightgcn}. Since $d\ll N$, therefore, the trainable parameters for CF-based models are in general much more than those of GNN models. The final embeddings of GNN models are output from the last layer. However, for the CF-based model, since the initial embeddings $\mathbf{E}^{(0)}$ are trainable and contain important structure information, existing methods all consider combining embeddings output from all layers. For example, NGCF concatenates the embeddings from $0$ layer to $L$ layers. LightGCN averages the embeddings of $L$ layers. In our paper, we use the same idea as in LightGCN that averages the embeddings from all layers. Hence, the final embeddings in graph CF models already employ the graph residual structure~\cite{Zhang2019GResNetGR} like in JKNet~\cite{xu18jk}.

\section{Propagation on Bipartite Graph}\label{sec:proof}
In this section, we introduce the information propagation over bipartite graphs and prove the existence of the oscillation problem. Intuitively, the oscillation is caused by the information of user nodes only propagates to item nodes, and vice versa. Before studying the oscillation problem on bipartite graphs, we present how the information propagates on an irreducible and aperiodic graph.
\begin{mydef}
    (Regular Graph): A regular graph is irreducible and aperiodic. Given an input graph $\mathcal{G}=\left(\mathcal{V},\mathcal{E}\right)$, $\mathcal{G}$ is irreducible iff for any two nodes $v_i$ and $v_j$, they are accessible to each other. Meanwhile, $\mathcal{G}$ is aperiodic iff it is non-bipartite.
\end{mydef}
\begin{lemma}~\label{lemma:converge}
Given an unweighted regular graph $\mathcal{G}$, if its corresponding matrix is asymmetric, starting from any initial distribution vector $\mathbf{x}_{0} \in \mathbb{R}^{n}$ with $\mathbf{x}_{0}\geq 0$ and $\|\mathbf{x}_0\|_1 =1$, the random walk propagation over the graph has a unique stationary distribution vector $\bm{\pi}$, i.e. $\lim_{t \rightarrow \infty} \Tilde{\mathbf{A}}^{t} \mathbf{x}_{0}=\bm{\pi}$ where $\bm{\pi}(i)=\frac{d\left(v_{i}\right)}{2|\mathcal{E}|}$. $d(v_i)$ represents the degree of node $v_i$. $|\mathcal{E}|$ denotes the total number of edges.
\end{lemma}
The proof of this Lemma is given in appendix.

\begin{mydef}
(Bipartite Graph): A bipartite user-item graph is defined as $\mathcal{B}=(\mathcal{U},  \mathcal{I},  \mathcal{E})$, where $\mathcal{U}$ and $\mathcal{I}$ are two disjoint sets of nodes, i.e., $\mathcal{U}\cap \mathcal{I}=\phi$, denoting users and items, respectively. Every edge $e\in \mathcal{E}$ has the form $e=(u,i)$, where $u\in \mathcal{U}$ and $i\in\mathcal{I}$. The corresponding adjacent matrix $\mathbf{A}=\{0,1\}^{(|\mathcal{U}| + |\mathcal{I}|) \times (|\mathcal{U}| + |\mathcal{I}|)}$ of the bipartite graph is defined as:
\begin{equation}\label{interaction matrix}
\mathbf{A} = \begin{bmatrix}
0 & \mathbf{R} \\
\mathbf{R}^\top & 0
\end{bmatrix},\quad \text{where}\quad
\mathbf{R}_{u,i} = \left\{\begin{matrix}
1 & \text{if $(u, i)\in \mathcal{E}$} \\
0 & \text{otherwise}.
\end{matrix}\right.
\end{equation}
\end{mydef}
If taking account of only user (item) nodes, and creating an edge between users (items) if they are connected by a common item (user), we construct a user (item) side graph of the original bipartite graph. The associated side graph of a bipartite graph is defined as:
\begin{mydef}
(Side Graph): Given a bipartite graph $\mathcal{B}=(\mathcal{U},  \mathcal{I},  \mathcal{E})$, a user side graph of it is defined as  $\mathcal{G}_{u}=(\mathcal{U},\mathcal{E}_{u})$, where an edge $e\in\mathcal{E}_{u}$ has the form $(u_1,u_2)$ and $u_1,u_2 \in\mathcal{U}$. The edge $(u_1,u_2)$ is constructed from original bipartite graph $\mathcal{B}$, where $\exists i\in\mathcal{I}$ s.t. $(u_1,i)\in\mathcal{E} \wedge (u_2,i) \in \mathcal{E}$. Similarly, we can define an item side graph of $\mathcal{B}$ as $\mathcal{G}_{i}=(\mathcal{I},\mathcal{E}_{i})$.
\end{mydef}

Lemma~\ref{lemma:converge} illustrates that the propagation on a regular graph has a stationary distribution. However, the propagation over a bipartite graph never converges to a stationary distribution. Consider $\mathbf{x}_0$ to be a distribution only on one side of the bi-partition, i.e., only on $\mathcal{U}$ or only on $\mathcal{I}$. Then, the distribution of $\mathbf{x}_{1}$ at the first step random walk moves entirely to the other side, and thus, by induction, $\mathbf{x}_t$ is different for $t$s with different parity. This phenomenon is \textit{oscillation}.
\begin{mydef}
(Graph Oscillation): Given an input graph $\mathcal{G}$, starting from any initial distribution, if the random walk over $\mathcal{G}$ has two stationary distributions respectively on even steps and odd steps, i.e., $\lim_{t\rightarrow\infty}\Tilde{\mathbf{A}}^{2t}\mathbf{x}_0=\bm{\pi}^*_1$, $\lim_{t\rightarrow\infty}\Tilde{\mathbf{A}}^{2t+1}\mathbf{x}_0=\bm{\pi}^*_2$, and $\bm{\pi}^*_1 \neq \bm{\pi}^*_2$, there is an oscillation problem associated with $\mathcal{G}$.
\end{mydef}

\begin{theorem}~\label{theorem:oscillation}
  Given a bipartite graph $\mathcal{B}=(\mathcal{U},  \mathcal{I},  \mathcal{E})$, if its associated user and item side graph are both regular graphs, there is an oscillation problem associated with $\mathcal{B}$. 
\end{theorem}


We provide the proof and the boundary of oscillation in the Appendix. Theorem~\ref{theorem:oscillation} indicates that the oscillation problem exists on bipartite graphs. This suggests that stacking many layers results in the oscillation problem. Later, we will show oscillation occurs as few as $4$ layers in experiments.

\begin{figure*}
    \centering
    \includegraphics[width=0.75\linewidth]{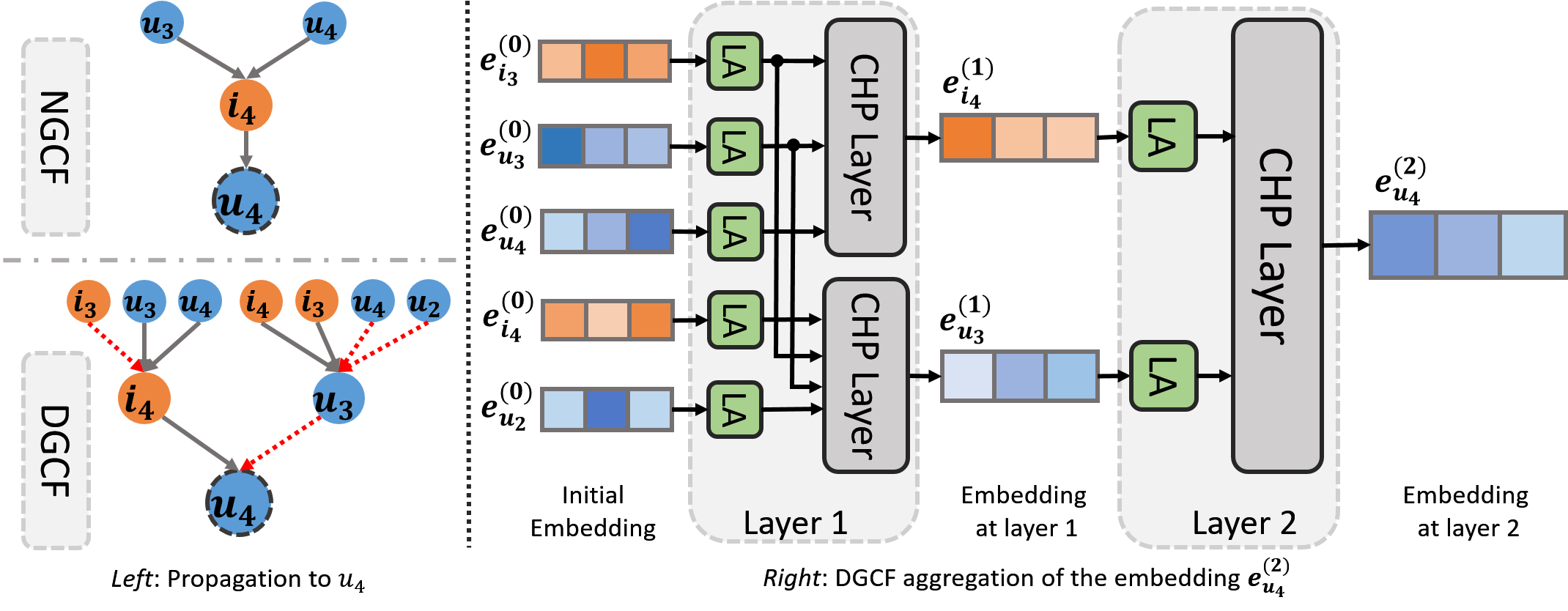}
    \vspace{-2mm}
    \caption{\textit{Left:} Comparison of NGCF and DGCF on the information propagation to $u_4$. Red dash lines are the cross-hop propagation. The original bipartite graph is in Figure~\ref{fig:example_osci}. \textit{Right:} The DGCF framework to learn embedding of user $u_4$ at layer $2$. Green `LA' blocks and gray `CHP layer' blocks are the  locality-adaptive layer and cross-hop propagation layer, respectively.}
    \label{fig:framwork}
\end{figure*}

\section{Proposed Model}
In this section, we present the structure of our proposed deep \textbf{D}eoscillated adaptive  \textbf{G}raph \textbf{C}ollaborative \textbf{F}iltering~(DGCF) model. DGCF is a multi-layer GNN model that propagates the CF signals over the bipartite graph. Instead of propagating information to direct neighbors, it has Cross-Hop Propagation~(CHP) layers that resolve the oscillation problem. Moreover, its Locality-Adaptive~(LA) layers learn the propagating factor for each node. Stacking multiple layers adaptively controls the information to propagate at different layers. The framework of aggregating embedding is present in Figure~\ref{fig:framwork}.

\subsection{Cross-Hop Propagation Layer}
CHP layer propagates the information for both direct and cross-hop neighbor nodes, which changes the bipartite structure to a regular graph. The additional cross-hop interactions are exactly from user and item side graphs of the bipartite graph. Thus, there exists only one stationary distribution on different parity, i.e., deoscillation. As illustrated on the left-hand side of Figure~\ref{fig:framwork}, the red dash lines represent additional cross-hop propagation of DGCF, compared with only direct propagation of NGCF model. The node embedding for user $u$ and item $i$ at $l$-th layer are ${\mathbf{e}_{u}^{(l)}}\in\mathbb{R}^{d}$ and $\mathbf{e}_{i}^{(l)}\in\mathbb{R}^{d}$, respectively. CHP layer outputs the embedding of each node by aggregating the information of neighbors, e.g., the embedding of a user $u$ at $l-$th layer:
\begin{equation}\label{eq:CHP layer}
\mathbf{e}_u^{(l)}= \sum_{j\in\Tilde{\mathcal{N}}_{u}}\alpha_{j}^{(l)}p_{j}\mathbf{e}_{j}^{(l-1)},
\end{equation}
where $\Tilde{\mathcal{N}}_{u}$ denotes neighbors, $p_j$ is the fixed normalizing factor, and ${\alpha_j^{(l)}}$ denotes the adaptive locality weight of node $j$. $\Tilde{\mathcal{N}}_{u}$ contains both direct and cross-hop neighbors of user $u$, as illustrated in the left part in Figure~\ref{fig:framwork}. For example, regarding the bipartite graph in Figure~\ref{fig:example_osci}, both $i_4$ and $u_3$ are included in $\Tilde{\mathcal{N}}_{u}$. $p_j$ denotes the importance of the neighbor $j$. It is a fixed value during training, e.g., the Laplacian normalizing coefficient. ${\alpha_j^{(l)}}$ are adaptively learned during training, which adjusts to the local structure. Analogously, we calculate the item embeddings by also using CHP layers. 

\subsection{Locality-Adaptive Layer}
DGCF has an LA layer before a CHP layer, which adaptively controls the propagation process for each node. It assigns a locality weight for each node at $l-$th layer, thus denoted as ${\alpha^{(l)}_{j}}$ as in Eq.~(\ref{eq:CHP layer}). Since the value of it should be from 0 to 1, we use sigmoid activation to train the weights, i.e.,
\begin{equation}
\bm{\alpha}^{(l)}=\sigma(\mathbf{w}^{(l)}_{LA}), \quad \text{where}\quad \mathbf{w}^{(l)}_{LA}\in\mathbb{R}^{|\mathcal{U}| + |\mathcal{I}|}.
\end{equation}
${\mathbf{w}^{(l)}_{LA}}$ is the trainable parameter vector for the $l-$th LA layer. $|\mathcal{U}|$ and $|\mathcal{I}|$ denote the number of users and items, respectively. Intuitively, LA layer assigns an influencing factor to all nodes before propagation, which is illustrated in Figure~\ref{fig:framwork}. Thus, during the aggregation process, it learns the importance of nodes. Since nodes exist in different local structures with varying densities, it turns out that the influence factor $\alpha$ controlling the propagation process should be adaptive to the locality. We verify the correlation between $\alpha$ and the locality in experiments.  

\textbf{Layer-wise adaptive manner.} Note that, as we learn different locality weights from layer to layer, the LA layers also adjust the propagation process in a layer-wise adaptive manner. This is a better way to propagate information on a graph. Each layer has distinct important substructures. Therefore, it reduces the redundancy compared with fixed propagation patterns. 

\subsection{Matrix Form}
In this section, we present the matrix form propagation of Eq.~(\ref{eq:CHP layer}) for all nodes. We use $\mathbf{E}^{(l)}\in\mathbb{R}^{(|\mathcal{U}|+|\mathcal{I}|)\times d}$ to denote embeddings for all user and item nodes at $l-$th layer. Before proceeding to the CHP layer, LA layer applies an importance coefficient to each node. Hence, the weighted embedding matrix is: 
\begin{equation}
    \Tilde{\mathbf{E}}^{(l)}=\bm{\Omega}^{(l)}\mathbf{E}^{(l)},
\end{equation}
where $\bm{\Omega}^{(l)}$ is diagonal and $\bm{\Omega}^{(l)} \in\mathbb{R}^{(|\mathcal{U}|+|\mathcal{I}|)\times(|\mathcal{U}|+|\mathcal{I}|)}$. 
${\bm{\Omega}^{(l)}}$ contains the importance factor $\bm{\alpha}^{(l)}$ in its elements, i.e., ${\bm{\Omega}^{(l)}}(j,j)=\bm{\alpha}^{(l)}_{j}$. The direct and cross-hop neighbors can be inferred by using the adjacent matrix $\mathbf{A}$ and the cross-hop matrix $\mathbf{C}=\mathbf{A}^{2}$, respectively. We adopt the Laplacian normalization~\cite{kipf17semi} of both matrices regarding the $p_i$ in Eq.~(\ref{eq:CHP layer}). The Laplacian matrix $\mathcal{L}$ and the cross-hop Laplacian matrix $\mathcal{L}_{c}$ are defined as:
\begin{equation}
    \mathcal{L} = \mathbf{D}^{-\frac{1}{2}}\mathbf{A}\mathbf{D}^{-\frac{1}{2}} \quad \text{and}\quad
    \mathcal{L}_{c} = \mathbf{D}_{c}^{-\frac{1}{2}}\mathbf{C}\mathbf{D}_{c}^{-\frac{1}{2}},
\end{equation}
respectively. $\mathbf{D}$ and $\mathbf{D}_c$ are both the diagonal degree matrices, by taking the row-sum of ${{\mathbf{A}}}$ and ${{\mathbf{C}}}$, respectively. Therefore, the final matrix form of the propagation is:
\begin{equation}
    \mathbf{E}^{(l)} = \left(\mathcal{L}+\mathcal{L}_{c}+\mathbf{I}\right)\Tilde{\mathbf{E}}^{(l-1)}.
\end{equation}
After $L-$layer propagation, we average the embeddings~\cite{he2020lightgcn} at each layer to construct the final embedding for prediction, i.e., \begin{equation}\label{eq:MATRIX_DGCF}
    \mathbf{E}^{*}=\frac{1}{L+1}\sum_{l=0}^{L}\mathbf{E}^{(l)}.
\end{equation} 

\subsection{High-pass Filtering}
The cross-hop matrix $\mathcal{L}_c$ propagates additional information. This also introduces extra computational costs. We observe that in large datasets, most~($>95\%$) of the entries in $\mathcal{L}_c$ are of very small values, which provides few informative signals for learning embeddings. Therefore, we filter the entries with values less than a threshold $\varepsilon$, i.e., using a filtered cross-hop matrix $\Tilde{\mathcal{L}}_c$ for graph convolution, where
\begin{equation}\label{interaction matrix}
\Tilde{\mathcal{L}}_{c}[i,j] = \left\{\begin{matrix}
\mathcal{L}_{c}[i,j] & \text{if $\mathcal{L}_{c}[i,j] > \varepsilon$} \\
0 & \text{otherwise}.
\end{matrix}\right.
\end{equation}

After \textit{high-pass filtering}, we only pass high values of the cross-hop matrix for graph convolution, which greatly improves the efficiency of the CHP layer on large datasets. We analyze how to select  optimal $\varepsilon$ in appendix.

\subsection{Optimization}
The initial embedding is generated from Xavier initializer~\cite{glorot2010understanding}. The final prediction between users and items is estimated by the inner product, i.e., $\hat{{y}}(u,i)=\mathbf{e}^{*\top}_u\mathbf{e}^{*}_i$. We use BPR~\cite{rendle2009bpr} loss to optimize the trainable weights:
\begin{equation}
    \mathcal{L} = -\sum_{(u,i,j)\in \mathcal{S}}\log\sigma\big(\hat{{y}}(u,i) - \hat{{y}}(u,j)\big) + \lambda\|\Theta\|_{2}^{2},
\end{equation}
where $\mathcal{S}$ is generated from the rule that we sample, for each user $u$, a positive item $i$ and a negative item $j$. The first term denotes the BPR interaction loss, and the second term is the regularization for parameters ($\lambda$ is a hyper-parameter).  We optimize the model by using the mini-batch Adam~\cite{kingma2014adam} optimizer in Tensorflow.

\subsection{Model Analysis}
The intuition of using \textbf{cross-hop} information to enhance the training process of GNN is also discussed in previous works~\cite{xu18jk,yang2018hop,abu2019mixhop,chen2020revisiting}. In this section, we compare DGCF model with existing models to clarify the difference. 

Taking residual model JKNet~\cite{xu18jk} as an  example, it adds additional skip connections from intermediate layers to the last layer of the GNN model. To simplify the comparison, we eliminate the convolutional weights and only keep the embedding propagation of JKNet~\footnote{Original JKNet also ignores the initial embeddings, which is rather important for RS.} as:
\begin{equation}\label{eq:jknet}
    \mathbf{E}_{JK}^{*} = \text{CONCAT}\{\mathbf{E}^{(0)}_{JK},\mathbf{E}^{(1)}_{JK},\dots,\mathbf{E}^{(L)}_{JK}\},
\end{equation}
where $\mathbf{E}_{JK}^{(l)}$ is the intermediate embeddings of JKNet and $\mathbf{E}_{JK}^{*}$ is the final output embeddings of JKNet. Since $\mathbf{E}_{JK}^{(l)}=\mathcal{L}\mathbf{E}_{JK}^{(l-1)}$, we can rewrite Eq.~(\ref{eq:jknet})  as:
\begin{equation}
    \mathbf{E}_{JK}^{*} = \text{CONCAT}\{\mathbf{E}^{(0)}_{JK},\mathcal{L}\mathbf{E}^{(0)}_{JK},\dots,\mathcal{L}^{L}\mathbf{E}^{(0)}_{JK}\}.
\end{equation}
Compared with DGCF in Eq.~(\ref{eq:MATRIX_DGCF}), which is rewritten as:
\begin{equation}
 \mathbf{E}^{*}=\frac{1}{L+1}\sum_{l=0}^{L}\left(\mathcal{L}+\mathcal{L}_{c}+\mathbf{I}\right)^l\Tilde{\mathbf{E}}^{(0)}.
\end{equation}
Other than the difference between \textit{concatenate} and \textit{average} operation, JKNet only considers $\mathcal{L}$ operators. It still suffers from the oscillation problem because $\mathcal{L}$ only propagates the information from one side graph to the other.

\begin{table*}[!h]
    \centering
    \caption{Overall Performance Comparison.}
    \begin{tabular}{lcccccccc}
    \hline
    \hline
    Dataset & \multicolumn{2}{c}{ML1M} &\multicolumn{2}{c}{Amazon}  &\multicolumn{2}{c}{Gowalla}  &\multicolumn{2}{c}{ML100K}  \\
    Metric@20 & Recall & NDCG & Recall & NDCG &Recall & NDCG & Recall & NDCG \\
    \hline
     BPR-MF &0.2653 &0.2149 &0.0762 &0.0588 &0.1371 &0.1126 &0.2894 &0.1907  \\
     GCN  &0.2628 &0.2086 &0.0701 &0.0569 &0.1426 &0.1161 &0.3025 &0.1919\\
     GCN+JK &0.2723 &0.2184 &0.0611 &0.0490 &0.1335 &0.1095 &0.3086 &0.1917 \\
     GC-MC &0.2611 &0.2069 &0.0578 &0.0475 &0.1181 &0.0967 &0.2966 &0.1883 \\
     NGCF &0.2693 &0.2164 &0.1117 &0.0886   &0.1485 &0.1196 &0.3146 &0.1978 \\
 LightGCN &0.2888 &0.2334 &0.1130 &0.0893 & 0.1584 &0.1309 & \underline{0.3399} &0.2137 \\
     \hline
    \textbf{DGCF} & \textbf{0.3075} & \textbf{0.2501} & \textbf{0.1351} & \textbf{0.1083} & \textbf{0.1707} & \textbf{0.1384} & \textbf{0.3536} & \textbf{0.2290}\\
    \text{DGCF-chp} & \underline{0.2975} & 0.2420 & \text{0.1241} & \text{0.0991} & \underline{0.1657} & \text{0.1344} & 0.3368 &  \text{0.2159}\\
    
     \text{DGCF-la} & 0.2967 & \underline{0.2425} & \underline{0.1289} & \underline{0.1030} & \text{0.1653} & \underline{0.1350} & \text{0.3385} &  \underline{0.2181}\\
    
    
     \hline
     \hline
    \end{tabular}
    \label{tab:main_results}
\end{table*}

Regarding the \textbf{complexity} of DGCF, as we mentioned before, all the initial embeddings are trainable in RS, which contains $dN$ parameters if given a total of $N$ user and item nodes. LA layer only introduces $LN$ parameters to train. Since $L\ll d$, e.g., $L=4$ and $d=64$ in our experiments, the model complexity is comparable to existing models. During implementation, we use the broadcasting mechanism in Tensorflow and thus simply use an LA vector to store the value rather than a matrix.

\section{Experiment}
In this section, we conduct extensive experiments to show the effectiveness of our proposed model on four benchmark datasets that include ML100K~\cite{harper2015movielens},  ML1M~\cite{harper2015movielens}, Amazon Movies and TV~\cite{he2016ups} and Gowalla~\cite{liang2016modeling}. All the datasets are split as training, validation, and testing data. In the experiments, we use the validation set to select the best model for testing. More details are in appendix. 

\subsection{Experimental Setting}
\begin{figure}
    \begin{subfigure}{0.24\textwidth}
     \centering
    \includegraphics[width=.9\textwidth]{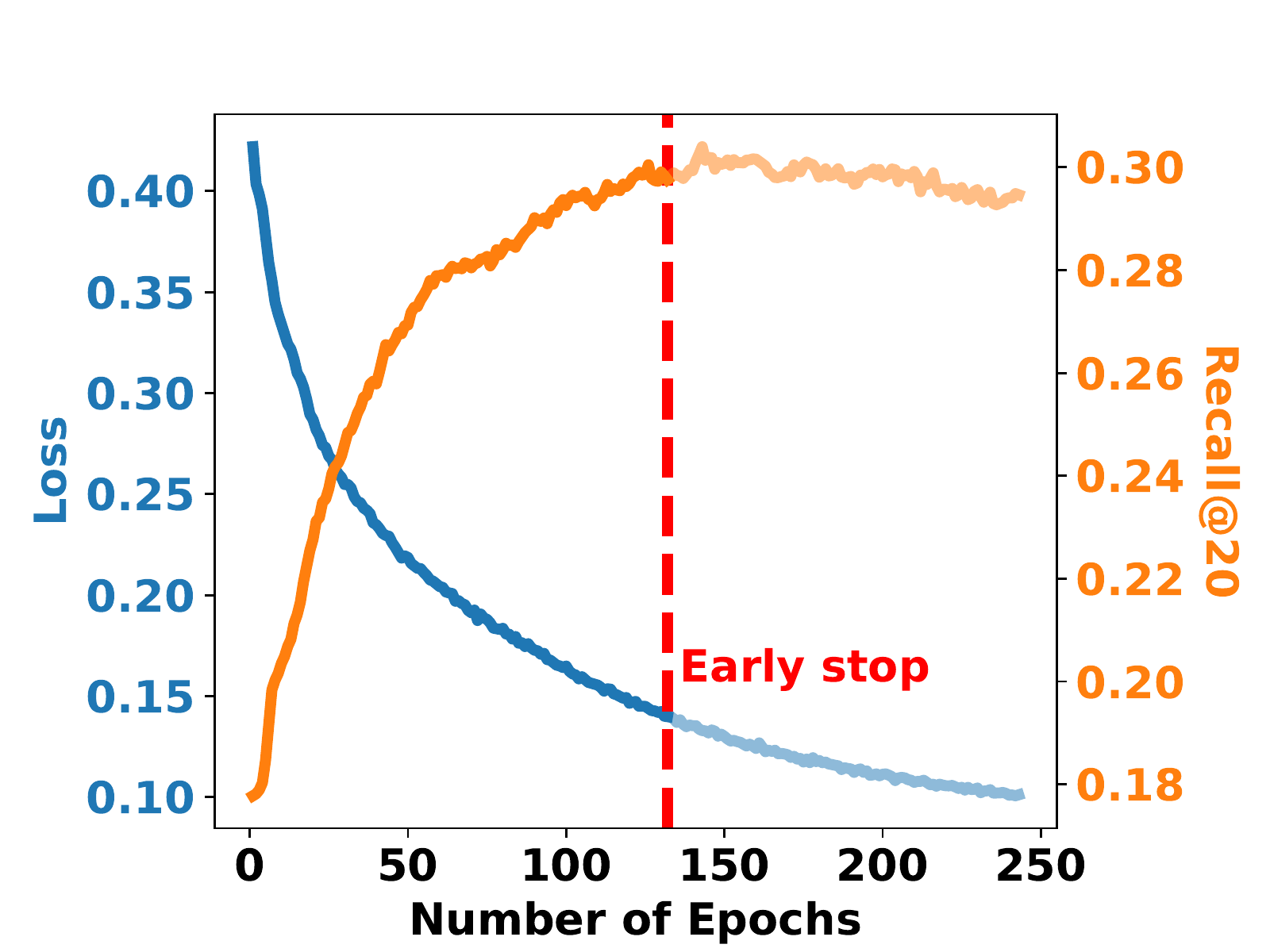}
    \caption{Loss and validation recall.}
    \label{fig:loss_validation}
    \end{subfigure}
    \hspace{-5mm}
    \begin{subfigure}{.24\textwidth}
     \centering
    \includegraphics[width=.9\textwidth]{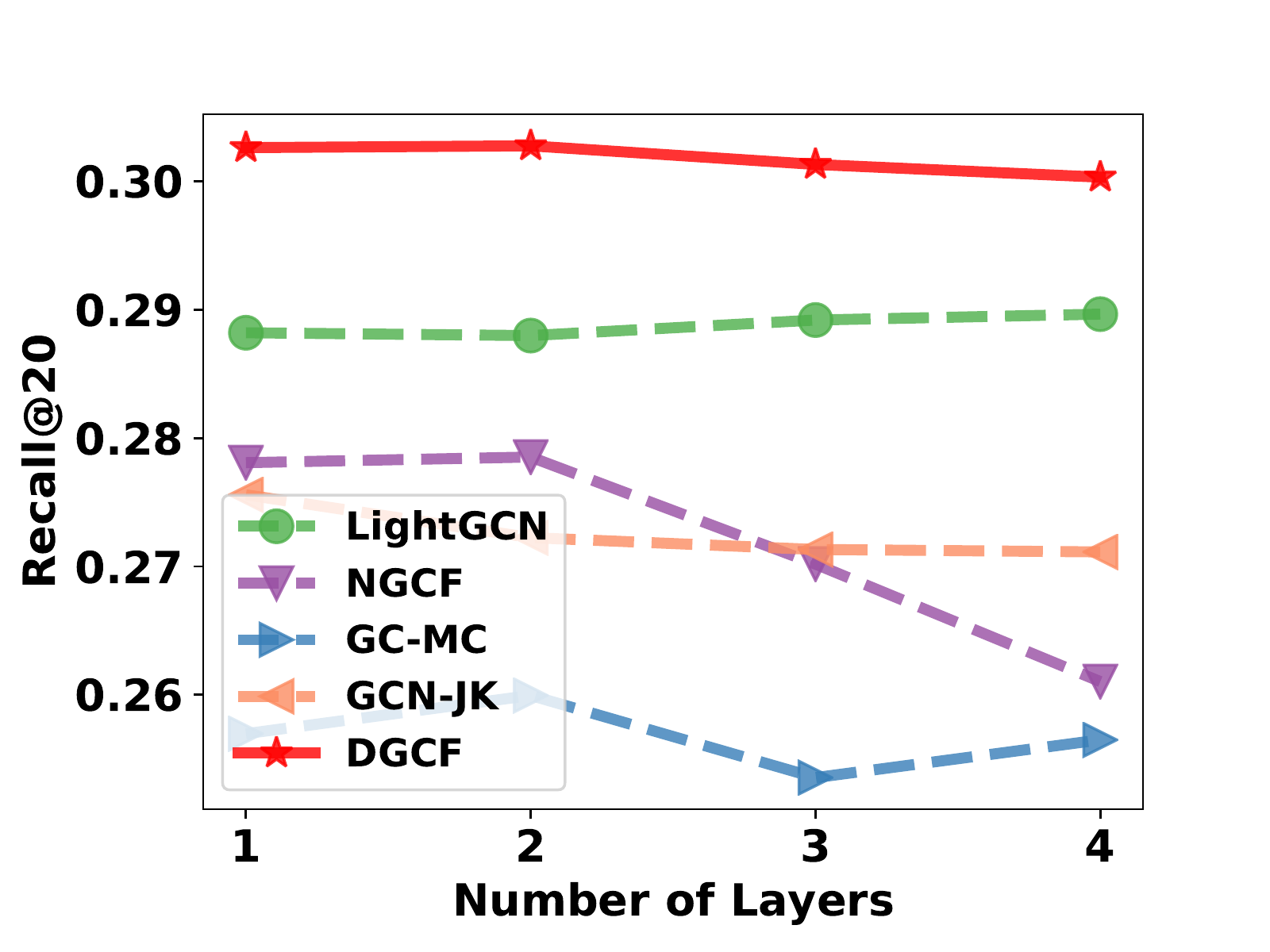}
    \caption{Recall w.r.t. \# of layers.}
    \label{fig:layer_ml1m}
    \end{subfigure}
    \vspace{-2mm}
    \caption{Training and tuning the number of layers of DGCF on ML1M dataset. Best viewed in colors.}
    \label{fig:training and tuning}
\end{figure}

To evaluate the model performance, several state-of-the-art methods are used for performance comparison, including BPR-MF~\cite{rendle2009bpr}, GCN~\cite{kipf17semi}, GCN with JKNet (GCN+JK)~\cite{xu18jk}, GC-MC~\cite{berg2017gcmc}, NGCF~\cite{wang19neural} and LightGCN~\cite{he2020lightgcn}. Besides DGCF, we also introduce DGCF-chp with only CHP layers and DGCF-la including only LA laysers as two variants into comparison. The evaluation metrics are Recall@K and NDCG@K, where K=20 by default.

All models are validated on the performance of Recall@20. The embedding size $d$ is searched from $\{32,64,128\}$ for all models. The learning rate for all models is searched from $\{10^{-5}, 10^{-4}, \dots, 10^{-1}\}$. The regularization factor $\lambda$ is searched from $\{10^{-5}, 10^{-4}, \dots, 10^{-1}\}$. We treat each interaction as a positive and pair it with a negative item that the user have no interactions with. The layer number $L$ (except for BPR-MF) is  searched from 1 to 4. The best hyper-parameters are searched based on the validation results. Early stop, i.e., the performance value on validation data not growing for $5$ times, is applied. The training process and the effects of the number of layers are in Figure~\ref{fig:training and tuning}. More implementing details, including the discussion on drop-edge and $\varepsilon$, are elaborated in Appendix.

\subsection{Overall Evaluation}

In this section, we present the overall comparing performance. The overall performance is shown in Table~\ref{tab:main_results}. The best performance values on different datasets are in bold and the second-best values are underlined. Our DGCF model outperforms others in all cases since it has CHP layer to remedy oscillation problem and LA layers that  adaptively learn influence factors for nodes.  We also create two variants for ablation study, with only CHP layers and only LA layers, i.e., DGCF-chp and DGCF-la, respectively. These two variants yield better values on almost all datasets compared with other baselines, which indicates the benefits of applying CHP and LA layers. But they are still worse than the default DGCF. Therefore, we should stack CHP layers and LA layers to overcome aforementioned problems.

For those baselines, GC-MC performs the worst compared with other models. The poor performance of GC-MC can be the result of the final MLP layer, which harms the effectiveness of structural regularization and thus overfits the training data~\cite{Liu2019JSCNJS}. GCN and GCN+JK have similar performance, while GCN gets slightly better results on large datasets (e.g., Amazon and Gowalla), which indicates directly applying graph residual structure cannot benefit the RS. Compared with other GCN models, LightGCN performs the best. The reason is that LightGCN removes redundant parameters (e.g., linear transformation) and non-linear activation. Moreover, by averaging the outputs of different layers, the expressiveness of it is enhanced by a large margin. However, LightGCN still suffers from the oscillation problem and is not able to adapt to the locality of nodes, which are solved by our DGCF model.


\begin{figure}
\centering
\includegraphics[width=0.45\textwidth]{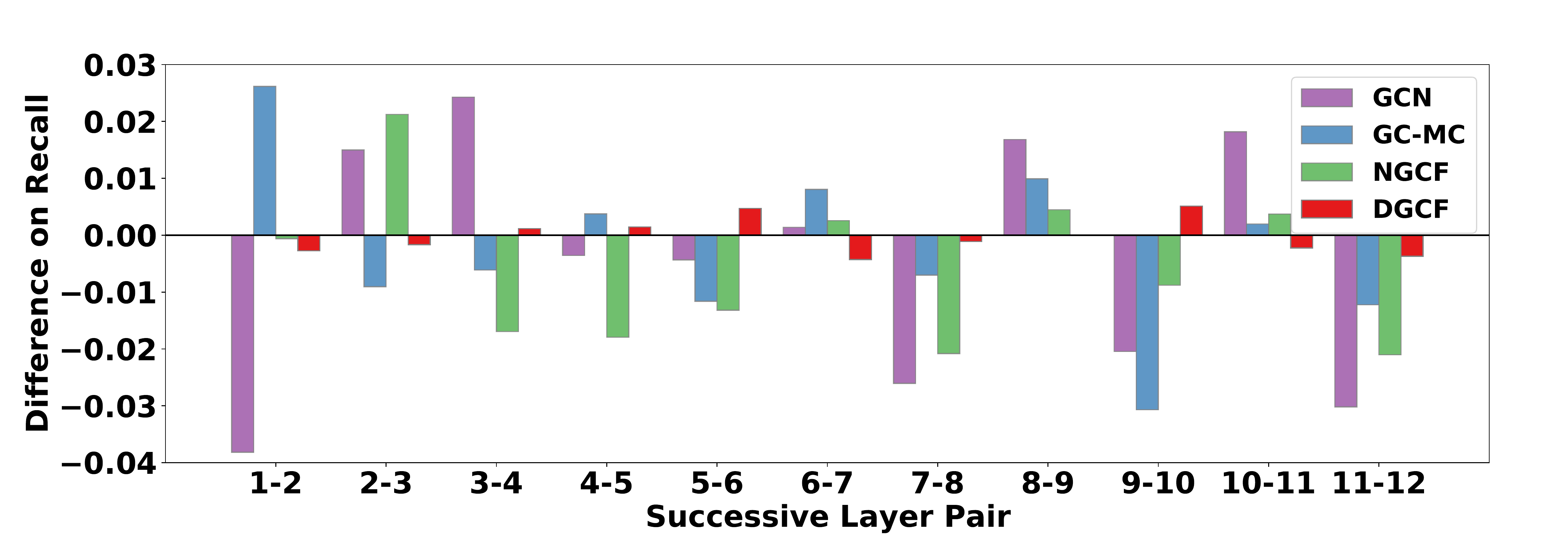}  
\vspace{-2mm}
\caption{Difference of Recall on successive layer pairs.}
\label{fig:diff}
\end{figure}

\subsection{Discussion on Oscillation Problem}

\begin{table}
    \centering
    \begin{tabular}{l|cccc}
    \hline
    Recall@20 & Layer 1 & Layer 2 & Layer 3& Layer 4  \\
     \hline
     GCN &0.2589 &0.2577 &0.2332 &0.2628  \\
     GCMC &0.2574 &0.2602 &0.2328 & {0.2646} \\
     NGCF &0.2107 &0.2111 &0.1767 &0.1759 \\
     LightGCN &{0.2930} & {0.2844} & {0.2543} & 0.1968 \\
     \hline
     DGCF & \textbf{0.3037} & \textbf{0.3041} & \underline{0.3027} & \underline{0.3012}\\
     DGCF-chp & \text{0.2961} & \text{0.2983} & \text{0.2943} & \text{0.3011}\\
     DGCF-la & \underline{0.2994} & \underline{0.3040} & \textbf{0.3031} & \textbf{0.3019}\\
         \hline
    \end{tabular}
     \caption{Performance Comparison on Single Layer.}
    \label{tab:Diff_layers}
 \end{table}

To understand the oscillation problem, we conduct experiments on ML100K. All models run 10 times for $1-12$ layers. We only use embeddings output from the final layer to predict. The differences of the average performance on Recall among successive layers are reported in Figure~\ref{fig:diff}. We observe that the performance goes up and down cyclically (i.e., oscillation), which matches the definition. Though oscillation theoretically appears with infinite layers, it occurs with a few layers in practice as the performance shows in Figure~\ref{fig:diff}. We notice that the oscillation amplitude of DGCF is the smallest one compared with other methods, which shows its ability of deoscillation. Additionally, we conduct experiments on ML1M by using only the embeddings at the last layer, which is illustrated in Table~\ref{tab:Diff_layers}. LightGCN has good performance in layer 1 and layer 2. But the performance drops quickly in layer 3 and layer 4. It is even worse than GCN and GCMC on layer 4. LightGCN only aggregates neighbor embeddings repeatedly for deep layers, which is vulnerable to both over-smoothness and oscillations. GCN performs better in layer 4 than in layer 1 and layer 2, which suggests it requires multiple layers. But it is observed that 3-layer is worse than 2 and 4, which results from oscillation. DGCF and its variants consistently yield good performance on all depths, which indicates its robustness. DGCF-la performs better on layer 3 and 4, which indicates the benefits of its adaptive ability. DGCF-chp also yields good performance on layer 3 and 4, which shows the deoscillation ability of CHP layers. 

\subsection{Discussion on Varying Locality}
\begin{figure}
\centering
\begin{subfigure}{.2\textwidth}
  \centering
  \includegraphics[width=1\textwidth]{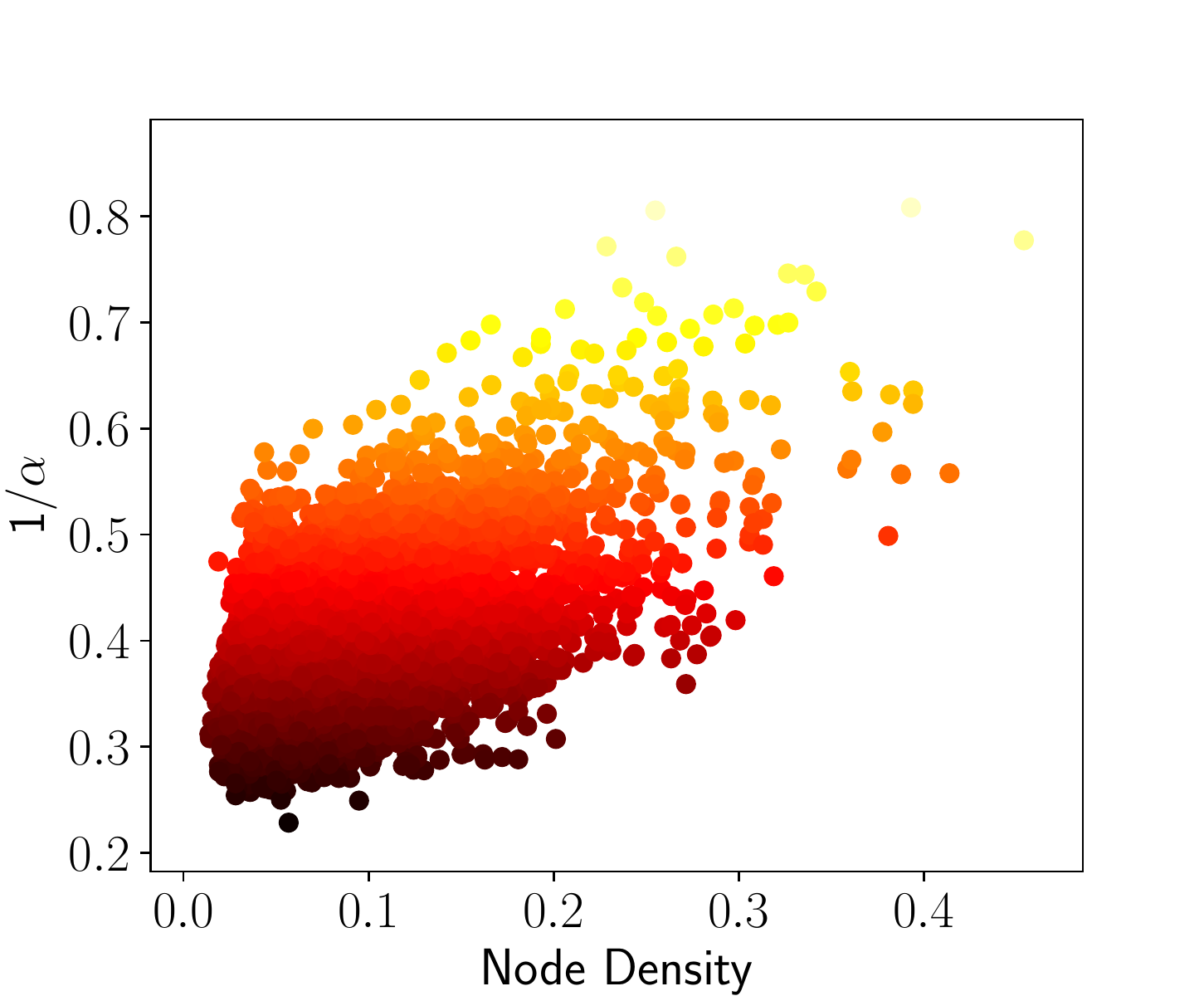} 
  \caption{Users. $\rho=0.578$}
  \label{fig:user_correlation}
\end{subfigure}
\begin{subfigure}{.2\textwidth}
  \centering
  \includegraphics[width=1\textwidth]{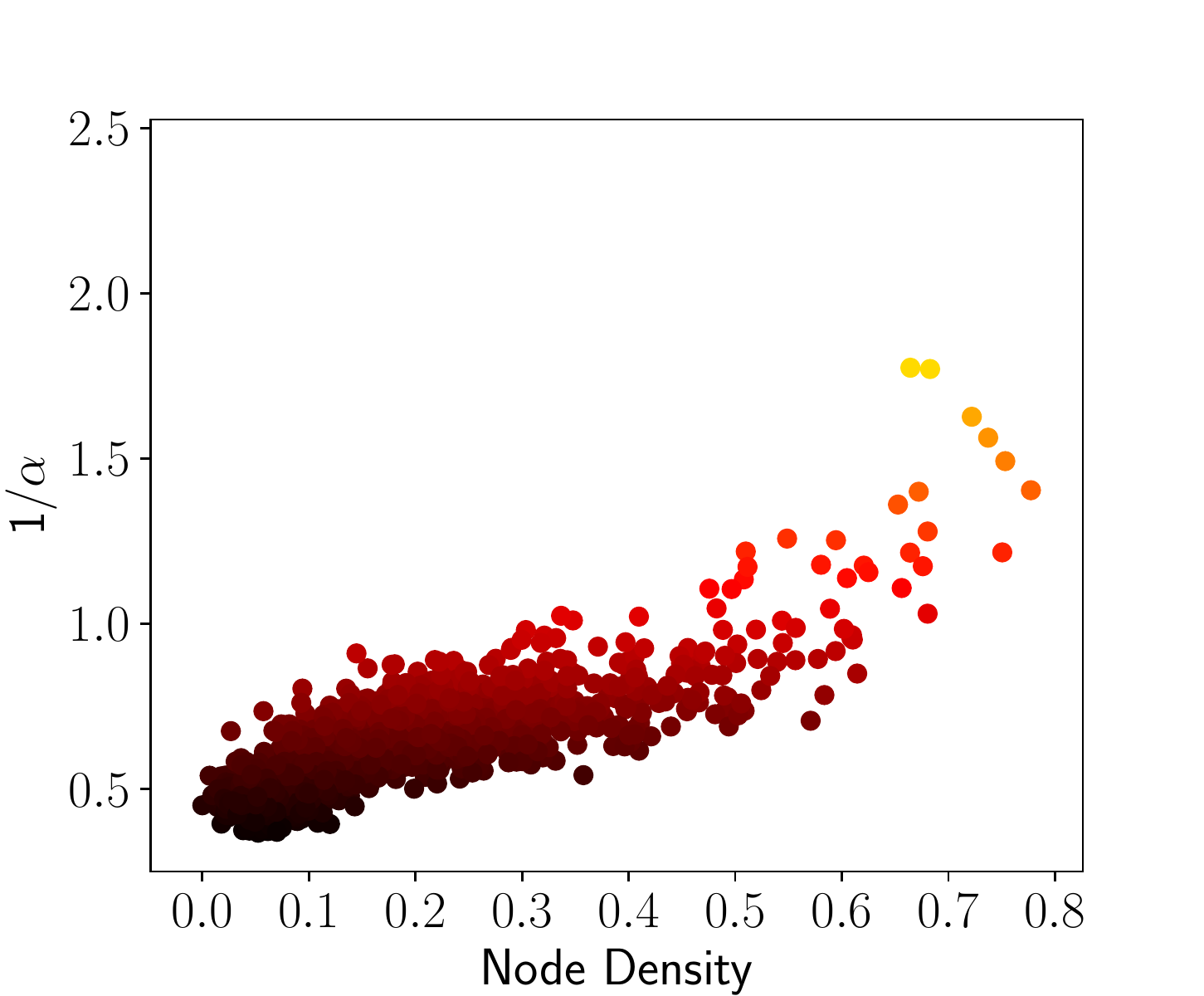}
  \caption{Items. $\rho=0.835$}
  \label{fig:item_correlation}
\end{subfigure}
\vspace{-2mm}
\caption{Correlation coefficient of LA layer and the density of nodes on ML1M. Left: users. Right: items. }\label{fig:correlation}
\end{figure}
In this section, we discuss the correlation between the learned influencing factor $\bm{\alpha}$ and the density of nodes. Intuitively, the LA layer balances the information propagation for different nodes, i.e., nodes with lower density should be assigned a higher influencing factor $\alpha$. We train DGCF on ML1M dataset. The distributions of nodes' density and the reciprocal value of $\alpha$ are illustrated in Figure~\ref{fig:correlation}. To better view the correlation, node's density is calculated as normalization of the logarithm of its degree. The correlation coefficients $\rho$ w.r.t. users and items are $0.578$ and $0.835$, respectively. It indicates $1/{\alpha}$ is positively correlated with the density of nodes, which proves that LA layer balances the influence of nodes regarding the varying locality problem. We also discuss in the appendix about the performance of nodes w.r.t. their density. 

\subsection{Discussion on Propagation Pattern}
\begin{figure}[t]
    \begin{subfigure}{0.11\textwidth}
  \centering
  \includegraphics[width=\linewidth]{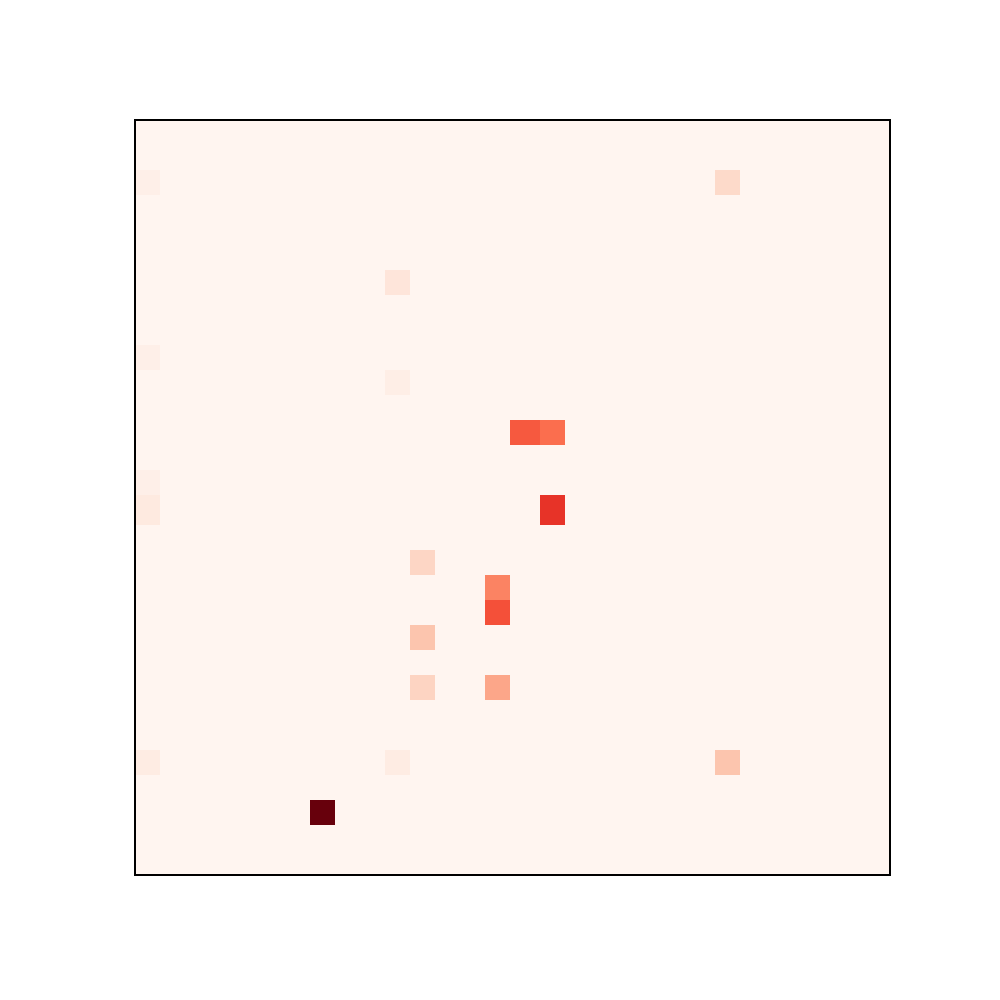}  
  \caption{Layer 1}
  \label{fig:sub-first}
\end{subfigure}
\begin{subfigure}{0.11\textwidth}
  \centering
  \includegraphics[width=\linewidth]{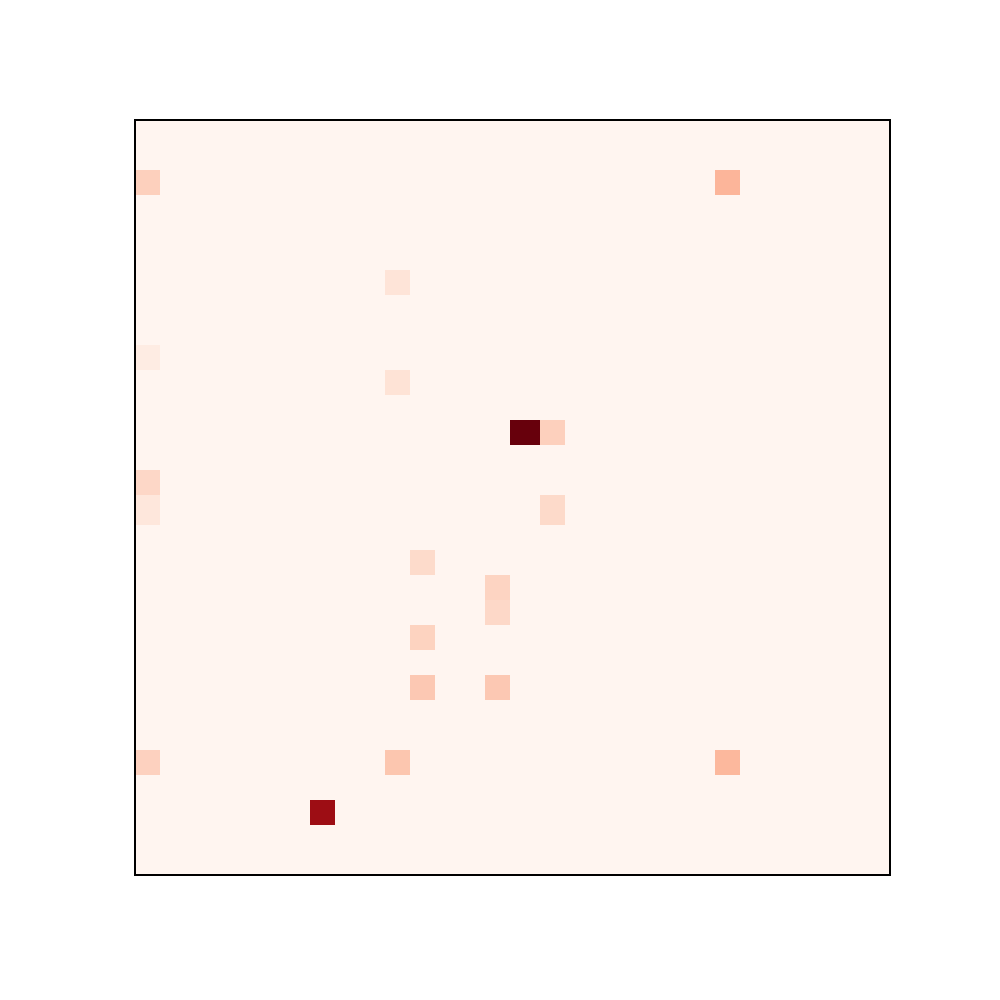}  
  \caption{Layer 2}
  \label{fig:sub-first}
\end{subfigure}
\begin{subfigure}{0.11\textwidth}
  \centering
  \includegraphics[width=\linewidth]{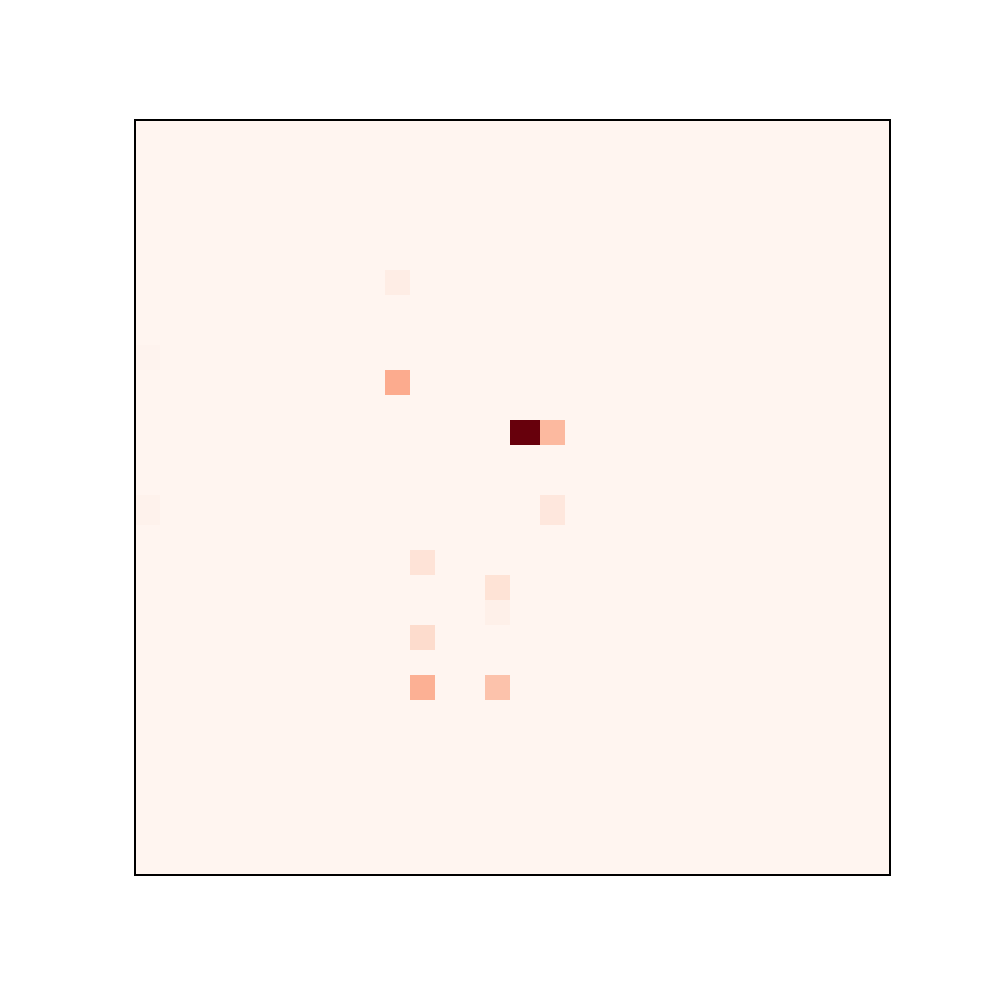}  
  \caption{Layer 3}
  \label{fig:sub-first}
\end{subfigure}
\begin{subfigure}{0.11\textwidth}
  \centering
  \includegraphics[width=\linewidth]{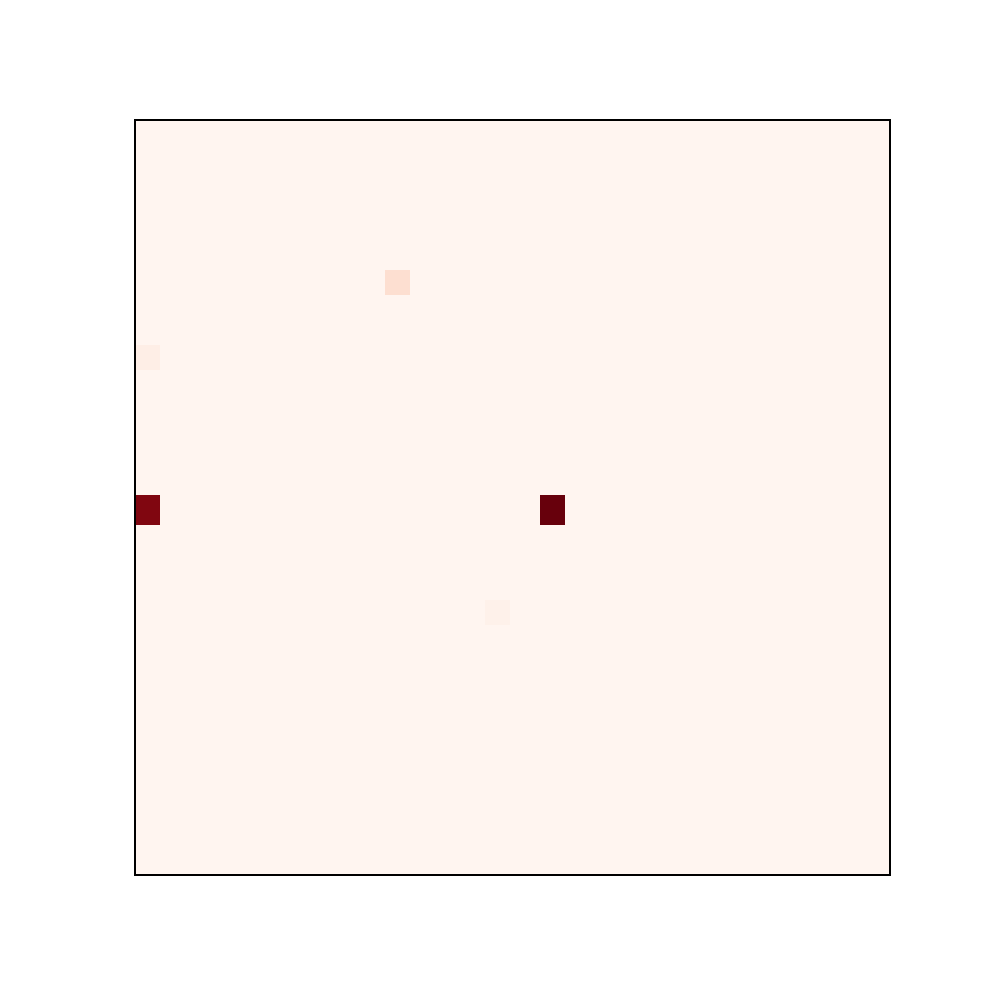}  
  \caption{Layer 4}
  \label{fig:sub-first}
\end{subfigure}
\vspace{-2mm}
\caption{Propagation pattern varies on different layers. Darker pixels are greater values. }\label{fig:prop_pattern}
\end{figure}

In this section, we conduct experiments to show how DGCF learns layer-wise aggregation patterns. Existing GNN RS models only have fixed propagation patterns. For example, NGCF only propagates information to direct neighbors at different layers. In contrast, DGCF has LA layer that assigns each node an influencing factor that controls the propagation pattern. Since DGCF has $L$ different LA layers, it has $L$ different propagation patterns. We implement a $4-$layer DGCF model and train it on ML1M data. We present the propagation patterns on each layer in Figure~\ref{fig:prop_pattern}. Due to the sparsity of the adjacency matrix, we zoom in a local patch (30$\times$30) to view the variations. Each pixel denotes the normalized value of the product of influence factor and Laplacian matrix, i.e., $\Omega(\mathcal{L}+\mathcal{L}_{c})$. We observe that the propagation pattern varies on different layers. On the first layer, it is a matrix containing almost every original edge, which suggests information is propagated out widely. Then, on layer 2, 3, and 4, the value on each edge varies and shrinks to a few important edges, which implies the propagation pattern adaptively concentrates on critical substructures.

\begin{figure}
    \centering
    \begin{subfigure}{0.22\textwidth}
    \centering
    \includegraphics[width=\linewidth]{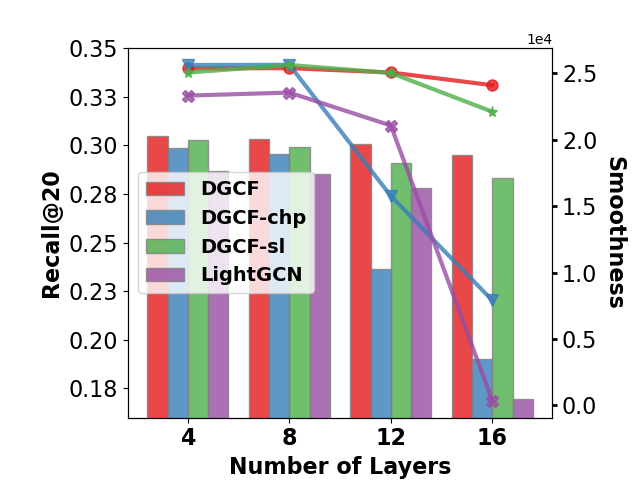}
    \caption{Recall@20}
    \end{subfigure}
    \hspace{-2mm}
    \begin{subfigure}{0.22\textwidth}
    \centering
    \includegraphics[width=\linewidth]{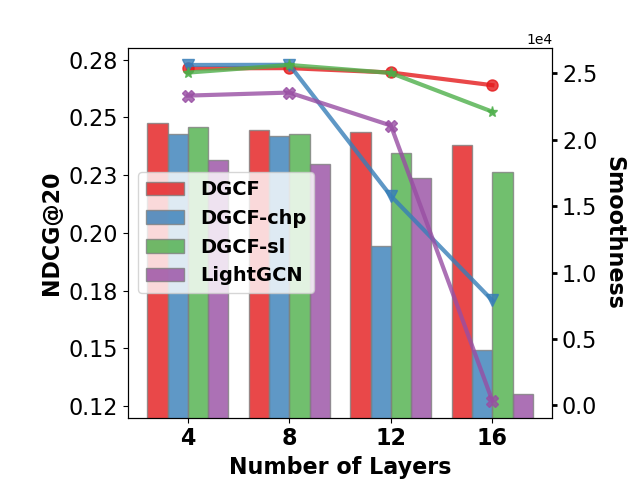}
    \caption{NDCG@20}  
    \end{subfigure}
    \vspace{-2mm}
    \caption{Performance and smoothness on deep layers. Lines represent the smoothness (lower is smoother). Bars are the performance, i.e. recall and NDCG.}
    \label{fig:deep_depth}
\end{figure}

\subsection{Variants Analysis}
In this section, we conduct the ablation study and discuss other variants of DGCF. The overall comparison of variant DGCF-chp and DGCF-la is already presented in Table~\ref{tab:main_results}. And the comparison on single layer is shown in Table~\ref{tab:Diff_layers}. Additionally, to study the effect of deep layers, we conduct experiments on ML1M dataset with layer number in $[4,8,12,16]$. We create a new variant that shares LA layers named DGCF-sl. The performance and smoothness~\cite{he2020lightgcn} comparison are presented in Figure~\ref{fig:deep_depth}. We observed that LightGCN and DGCF-chp drop dramatically with deep layers, which is the result of over-smoothing (i.e. low smoothness). The reason is that they both have fixed propagation patterns and thus propagate redundant information. DGCF and DGCF-sl both perform better, though drop with deep depth. Additionally, compared with DGCF-sl, the better performance of DGCF shows the effectiveness of its layer-wise adaptive manner.




\section{Conclusion}

In this paper, we study the oscillation problem, varying locality problem, and fixed aggregation pattern when applying multi-layer GNN on bipartite graphs. We formally define the graph oscillation problem and prove its existence on bipartite graphs. To tackle the problems, we propose a new model DGCF, which stacks multiple LA layers and CHP layers. 
The overall experiments on four real-world datasets prove the effectiveness of DGCF. Moreover, we conduct detailed analysis experiments. The experiment on the depth of the model implies the oscillation on existing models occurs while DGCF has the smallest amplitude. Also, the experiment regarding the varying locality indicates that DGCF adaptively learns the influence factor correlated with the density. Additionally, the experiment on the aggregation pattern suggests that DGCF automatically adjusts the aggregation pattern in a layer-wise manner.

\bibliography{icml_ref}
\bibliographystyle{icml2021}

\appendix
\section{Theoretical Proof}

\subsection{Proof of Lemma 1}
\begin{proof}
Assume $\bm{\pi}$ is a stationary distribution vector such that $\Tilde{\mathbf{A}}\bm{\pi}=\bm{\pi}$. According to the definition of $\Tilde{\mathbf{A}}$ and the matrix multiplication rule, $\bm{\pi}(i)$ can be represented as
\begin{equation}
    \sum_j\Tilde{\mathbf{A}}(i,j)\bm{\pi}(j) = \sum_{j}\frac{w_{i,j}}{d(j)}\bm{\pi}=\bm{\pi}(i),
\end{equation}
where $w_{i,j}$ denotes the initial weight between node $v_i$ and $v_j$ and $d(i)$ denotes the rough degree that sums the weight of edges connected to the nodes in the graph, i.e., $d(i) = \sum_j w_{i,j}$. Therefore, we can obtain $\bm{\pi_j} \propto d_j$ since $d(i)$ respects to $w_{i,j}$. Specifically, we can set 
\begin{equation}
    \bm{\pi}(j)=\frac{d(j)}{\sum_k d(k)} = \frac{d(j)}{2|\mathcal{E}|}
\end{equation}
Thus, 
\begin{equation}
\sum_j\Tilde{\mathbf{A}}(i,j)\bm{\pi}(j) = \sum_j\frac{w_{i,j}}{d(j)}\frac{d(j)}{2|\mathcal{E}|} = \sum_j\frac{w_{i,j}}{2|\mathcal{E}|} = \frac{d(i)}{2|\mathcal{E}|}
\end{equation}
holds for any regular graph (note that $d(i)$ equals to $d(v_i)$).

\end{proof}

\subsection{Proof of Theorem 1}

\begin{proof}
Given a bipartite graph $\mathcal{B} = \{\mathcal{U}, \mathcal{I}, \mathcal{E}\}$, where $\mathcal{U}$ and $\mathcal{I}$ represent two groups of nodes and $\mathcal{E}$ represents the corresponding edges between the nodes. The associated normalized adjacent matrix is:
Given a bipartite graph $\mathcal{B} = \{\mathcal{U}, \mathcal{I}, \mathcal{E}\}$, where $\mathcal{U}$ and $\mathcal{I}$ represent two groups of nodes and $\mathcal{E}$ represents the corresponding edges between the nodes. The associated normalized adjacent matrix is:
\begin{equation}\label{eq:norm_A}
    \mathbf{\hat{A}}=\left[\begin{array}{cc}
{0} & {\mathbf{R}} \\
{\mathbf{R}^{\top}} & {0}
\end{array}\right],
\end{equation}
where $\mathbf{R}\in \mathbb{R}^{|\mathcal{U}|\times|\mathcal{I}|}$ is the normalized link matrix from node set $\mathcal{U}$ to node set $\mathcal{I}$.
For simple analysis, we assume the initial probability distribution on the nodes is $\mathbf{x}^{(0)} = [\underbrace{{x}_{u_{1}}^{(0)},{x}_{u_{2}}^{(0)},...,{x}_{u_{|\mathcal{U}|}}^{(0)}}_{\mathbf{x}_u^{(0)\top}},\underbrace{{x}_{i_{1}}^{(0)},\dots,{x}_{i_{|\mathcal{I}|}}^{(0)}}_{\mathbf{x}_i^{(0)\top}}]^\top$ for all the associated nodes, which is the normalized node feature. We apply one step of random walk on the graph $\mathcal{B}$, which is:
\begin{equation}
    \begin{split}
        \mathbf{x}^{(1)} &= \mathbf{\hat{A}}\mathbf{x}^{(0)}
        = \begin{bmatrix}
            \mathbf{R}\mathbf{x}_i^{(0)}\\
            \mathbf{R}^{\top}\mathbf{x}_u^{(0)}
        \end{bmatrix}.
    \end{split}
\end{equation}
And two steps random walk is:
\begin{align}
    \mathbf{x}^{(2)} = \mathbf{\hat{A}}\mathbf{x}^{(1)}
        = \begin{bmatrix}
      \mathbf{R}\mathbf{R}^{\top}\mathbf{x}_u^{(0)}\\
     \mathbf{R}^\top\mathbf{R}\mathbf{x}_i^{(0)}
    \end{bmatrix}.
\end{align}
Let $\mathbf{V}=\mathbf{R}\mathbf{R}^{\top}$ and $\mathbf{T}=\mathbf{R}^{\top}\mathbf{R}$, and if we continue the random walks for $2k$ steps, we have
\begin{equation}
  \begin{array}{ll}
\mathbf{x}^{(2k)} &= \mathbf{\hat{A}}\mathbf{\hat{A}}\mathbf{x}^{2(k-1)} \\
&= \begin{bmatrix}
    0 & \mathbf{R} \\
    \mathbf{R}^\top & 0 
\end{bmatrix}\begin{bmatrix}
    0 & \mathbf{R} \\
    \mathbf{R}^\top & 0
\end{bmatrix}\mathbf{x}^{2(k-1)} \\
 &= \begin{bmatrix}
    \mathbf{V} & 0 \\
    0 & \mathbf{T}
\end{bmatrix}\mathbf{x}^{2(k-1)}\\
&= \begin{bmatrix}~\label{eq:even_prob}
    \mathbf{V}^{k}\mathbf{x}_u^{(0)} \\
    \mathbf{T}^{k}\mathbf{x}_i^{(0)}
\end{bmatrix}.
\end{array}
\end{equation}
We can find that $\mathbf{V}\in\mathbb{R}^{|\mathcal{U}|\times |\mathcal{U}|}$ is the corresponding normalized adjacent matrix of user side graph. $\mathbf{T}\in\mathbb{R}^{|\mathcal{I}|\times |\mathcal{I}|}$ is the corresponding item side graph. Since $\mathbf{V}$ and $\mathbf{T}$ are both associated with regular graphs, when $k \rightarrow \infty$, the final probability is the stationary distribution $\bm{\pi}$. Hence, we have:
\begin{equation}\label{eq:even_convergence}
 \lim_{k \rightarrow \infty}\mathbf{x}^{(2k)} = \bm{\pi} = \begin{bmatrix}
            \bm{\pi}_{u} \\
            \bm{\pi}_{i}
        \end{bmatrix},
\end{equation}
where $\bm{\pi}_{u}$ and $\bm{\pi}_{i}$ are the stationary probability for user side graph and item side graph, respectively. Similarly, for the random walks with $2k+1$ steps, we have the 
\begin{equation}\label{eq:odd_convergence}
\begin{array}{ll}
      \lim\limits_{k \rightarrow \infty}\mathbf{x}^{(2k+1)} &= \lim\limits_{k \rightarrow \infty}\mathbf{\hat{A}}\mathbf{x}^{2k} = \mathbf{\hat{A}}\bm{\pi} \\
     &  = \begin{bmatrix}
            0 & \mathbf{R} \\
            \mathbf{R}^\top & 0
        \end{bmatrix}\begin{bmatrix}
            \bm{\pi}_{u} \\
            \bm{\pi}_{i}
        \end{bmatrix}   =\begin{bmatrix}
            \mathbf{R}\bm{\pi}_{i} \\
            \mathbf{R}^\top\bm{\pi}_{u}
        \end{bmatrix}=\bm{\pi}^{\prime}. 
\end{array}
\end{equation}
From Eq.~(\ref{eq:even_convergence}) and Eq.~(\ref{eq:odd_convergence}), we know that multi-step propagation on a bipartite graph would reduce the representation of the node from a column normalized feature to two different stationary distributions w.r.t. the parity of the number of steps. It suggests that the oscillation problem is a result of the parity of the number of layers of a multi-layer GNN model. Note the oscillation vanishes if and only if $\mathbf{R}\bm{\pi}_{i}=\bm{\pi}_{u}$.
\end{proof}

\begin{table*}[!ht]
	\caption{Datasets statistics}
	\label{tab:datastat}
	\centering
	\begin{tabular}{lcccccc}
        \hline		 
        \multirow{2}{*}{Dataset} &\multirow{2}{*}{\#Users} &\multirow{2}{*}{\#Items}
        &\multirow{2}{*}{Density}  &\multicolumn{3}{c}{\#Interactions}  \\
        \cline{5-7}
      &&& &Train &Validation &Test \\
		\hline                    
		MovieLens 100k &779 &1,169 &0.02285 &14,585 &2,068 &4,152 \\
	    MovieLens 1M &4,627 &1,840 &0.02490 &148,299 &21,362 &42,347 \\
		Amazon Movies\&TV   &14,432 &28,242 &0.00155  &441,771 &63,654 &126,085 \\
		Gowalla    &29,858 &40,981 &0.00084 &718,418 &103,862 &205,090 \\
		\hline
	\end{tabular}
\end{table*}

\subsection{Boundary of Oscillation}
In addition to the layer-parity-related stationary distribution, we further analyze whether there exists a bound between successive layers regarding the distribution. The bound is a theoretical proof that the final representation learned from the GCN model oscillates within a shallow range. Hence, oscillation would not affect too much on the tuning process of the GCN model w.r.t. the number of layers. Assuming that we are at even step which denoted as $2k$-th step, from Eq.~(\ref{eq:even_prob}), we have the following difference of the probability between $2k$-th step and $2k+1$-th step as:
\begin{equation}
    \begin{array}{ll}
\left\|\mathbf{x}^{(2k)}-\mathbf{x}^{(2k+1)}\right\|_{1} = \left\|
    \begin{bmatrix}
    \mathbf{V}^{k}\mathbf{x}_u^{(0)} \\
    \mathbf{T}^{k}\mathbf{x}_i^{(0)}
\end{bmatrix}
- \begin{bmatrix}
    \mathbf{R}\mathbf{T}^{k}\mathbf{x}_i^{(0)}\\
    \mathbf{R}^\top\mathbf{V}^{k}\mathbf{x}_u^{(0)}
\end{bmatrix}\right\|_{1},
\end{array}
\end{equation}
where the $\|\cdot\|_1$ is the 1-norm of the vector. Since we have $\mathbf{R}\mathbf{T}^{k} = \mathbf{V}^{k}\mathbf{R}$ and $\mathbf{R}^\top\mathbf{V}^{k}=\mathbf{T}^{k}\mathbf{R}^\top$, hence we have 
\begin{equation}\label{eq:difference_vt}
    \begin{array}{ll}
\left\|\mathbf{x}^{(2k)}-\mathbf{x}^{(2k+1)}\right\|_{1} = \bigg\|\begin{bmatrix}
    \mathbf{V}^{k}\big(\mathbf{R}\mathbf{x}_i^{(0)}-\mathbf{x}_u^{(0)}\big)\\
    \mathbf{T}^{k}\big(\mathbf{R}^\top\mathbf{x}_u^{(0)}-\mathbf{x}_i^{(0)}\big)
\end{bmatrix}\bigg\|_{1}
\end{array}.
\end{equation}
Since $\mathbf{R}$ $\mathbf{x}_i^{(0)}$ and $\mathbf{x}_u^{(0)}$ are constant given as the initial value of the graph, we denote $\mathbf{R}\mathbf{x}_i^{(0)}-\mathbf{x}_u^{(0)}$ and $\mathbf{R}^\top\mathbf{x}_u^{(0)}-\mathbf{x}_i^{(0)}$ as $\mathbf{\varepsilon}$ and $\mathbf{\xi}$, respectively. Then, the Eq.~(\ref{eq:difference_vt}) becomes:
\begin{equation}\label{eq:difference_leq}
    \begin{array}{ll}
\left\|\mathbf{x}^{(2k)}-\mathbf{x}^{(2k+1)}\right\|_{1} &= \left\|\begin{bmatrix}
    \mathbf{V}^{k} & 0 \\
    0 & \mathbf{T}^{k}
\end{bmatrix}\begin{bmatrix}
 \varepsilon\\
 \xi
\end{bmatrix}\right\|_{1} \\
& \leq \bigg\|\begin{bmatrix}
    \mathbf{V}^{k} & 0 \\
    0 & \mathbf{T}^{k}
\end{bmatrix}\bigg\|_{1} \cdot \bigg\|\begin{bmatrix}
 \varepsilon\\
 \xi
\end{bmatrix}\bigg\|_{1}
\end{array}.
\end{equation}
Since the 1-norm of the matrix is the maximum column sum~\cite{noble1988applied}, the final inequality in Eq.~(\ref{eq:difference_leq}) is:
\begin{equation}
\begin{array}{ll}
    & \left\|\mathbf{x}^{(2k)}-\mathbf{x}^{(2k+1)}\right\|_{1} 
     \leq \\
     &\max\bigg\{\max\limits_{1 \leq j \leq |\mathcal{U}|} \sum\limits_{i=1}^{|\mathcal{U}|}\left|v_{i j}\right|, \max\limits_{1 \leq j \leq |\mathcal{I}|} \sum\limits_{i=1}^{|\mathcal{I}|}\left|t_{i j}\right|\bigg\} \cdot \bigg\|\begin{bmatrix}
 \varepsilon\\
 \xi
\end{bmatrix}\bigg\|_{1},
\end{array}
\end{equation}
where the $v_{ij}$ and $t_{ij}$ is the corresponding entries of the matrix $\mathbf{V}^{k}$ and matrix $\mathbf{T}^{k}$ respectively. Here, we prove that the difference between the the odd steps and the even steps is bounded by the maximum value in the corresponding matrix. And since the maximum value in the matrix is always less than $1$ due to probability property, we can safely increase the layer number while the  amplitude will not exceed the bound. 

\section{Experimental Details}
\subsection{Dataset Description}\label{sec:dataset}
To evaluate the effectiveness of our proposed model, we conduct experiments on four public benchmark datasets: MovieLens 100k, MovieLens 1M, Amazon Movies and TV and Gowalla. After pre-processing, we summarize the statistics of three datasets in  Table~\ref{tab:datastat}. 
\begin{itemize}
\item \textbf{MovieLens 100K}: This is a commonly used benchmark dataset~\cite{harper2015movielens}. For this dataset, we maintain users with at least 5 interactions.
    \item \textbf{MovieLens 1M}: This movie rating dataset~\cite{harper2015movielens} has been widely used in the recommendation scenarios. To ensure the quality, we only retain ratings equal to 5 and apply a 10-core setting adopted in~\cite{wang19neural}, retaining the users and items with at least 10 interactions. Thus, we get 4,627 users and 1,840 items as well as 212,008 interactions.
\item \textbf{Amazon Movies and TV}: Amazon Movies and TV is another widely used dataset~\cite{he2016ups} for product recommendation. Here, we also hold ratings equal to 5 and adopt a 10-core setting on users and items, which results in 14,432 users and 28,242 items with 631,510 interactions.
    \item \textbf{Gowalla}: This is a checkin dataset~\cite{liang2016modeling} obtained from Gowalla, where the users share locations by checkins. After filtered by 10-core setting, the dataset contains 1,027,370 interactions between 29,858 users and 40,981 items. 
\end{itemize}
For each dataset, we randomly choose 70\% interactions as the training set, 10\% as the validation set, and the rest 20\% as the testing set. For each interaction in the training set, we treat it as a positive interaction and randomly select one negative interaction that is not in the training set.

In this section, we introduce the details of the implementation. For the overall evaluation, we first train the model on the training data. During training, we validate all models based on the performance of Recall@20 on validation data. We also use early-stopping to prevent models from overfitting if it reaches the highest value without growing for $5$ times. The best models are selected based on their performance on validation data. Finally, we test the selected models on testing data. For all models, we use the Xavier initializer for trainable parameters. We also randomly dropout partial edges to prevent oversmoothing.

\subsection{Hyperparameters}\label{sec:imp_detail}

Hyper-parameters of DGCF are chosen as follows: For ML1M dataset, we use 2-layer DGCF. The learning rate is $0.001$ and the regularization factor $\lambda=0.01$. The embedding dimension is 64. For Movies and TV dataset, we use 4-layer DGCF. The learning rate is $0.0001$ and the regularization factor $\lambda=0.01$. The embedding dimension is 128. For Gowalla dataset, we use a 3-layer structure DGCF. The learning rate is $0.0001$ and $\lambda=0.001$. The embedding dimension is 128. For ML100K, we use 3-layer DGCF. The learning rate is $0.001$ and the regularization factor $\lambda=0.01$. The embedding dimension is 128. Moreover, due to large size of cross-hop adjacency matrices of Movies and TV dataset and Gowalla dataset, we apply filtering technique to drop cross-hop weights less than $\varepsilon$, i.e, filtering $\mathcal{L}^{2}_{c}[{i,j}]<\varepsilon$, where $\varepsilon = 0.006$, $\varepsilon = 0.001$, $\varepsilon = 0.003$ and $\varepsilon = 0.004$ for ML100K, ML1M, Movies and TV and Gowalla datasets, respectively. We also employ the drop-edge~\cite{rong2019dropedge} operation during training to improve the robustness of DGCF. The dropping ratio is set to $0.1$ on all datasets, which will be discussed later. 

For other baseline methods, we set the learning rate to $0.001$ and $\lambda=0.00001$ for both ML100k and ML1M for all baselines. For Amazon Movies and TV, we set learning rate as $0.001$ and $\lambda=0.00001$. For Gowalla, the learning rate is $0.0001$ and $\lambda=0.001$. To make a fair comparison, we also use the same embedding size as DGCF. We choose the best score obtained from one layer to four layers for each GCN-based baseline. All baselines are reproduced and implemented based on code\footnote{\url{https://github.com/xiangwang1223/neural\_graph\_collaborative\_filtering}}.

\subsection{Implementation}
Our model is built upon tensorflow-gpu framework. The GPU machine we use is GeForce GTX 1080Ti. The CPU is Intel(R) Xeon(R) CPU E5-2620 v3 @ 2.40GHz. The memory size is 64 GB in total. On average, the running times of the 4-layer DGCF are 3.4 s/epoch, 22.0 s/epoch, 62.1 s/epoch, and 100.6 s/epoch on ML100K dataset, ML1M dataset, Movies and TV dataset, and Gowalla dataset, respectively.



\begin{figure}
      \centering
    \includegraphics[width=.8\columnwidth]{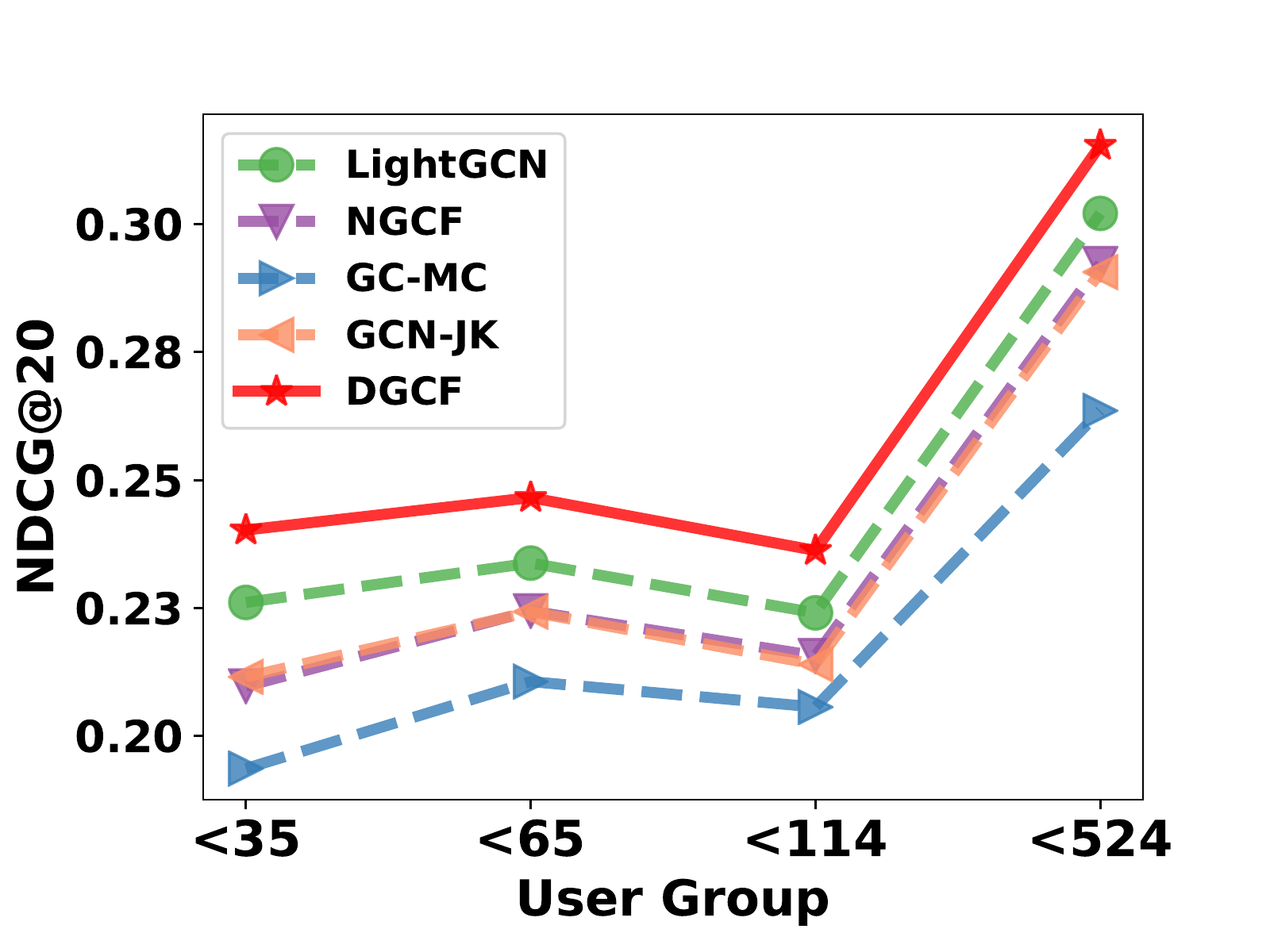}
    \caption{Performance w.r.t. different density groups of users on ML1M dataset.}
    \label{fig:density_ml1m}
\end{figure}
\subsection{Comparison on Density Levels}
In this section, we discuss the model performance under different users' sparsity levels based on ML1M dataset. We split test users into four groups according to the number of interactions they have, and each group has same number of interactions. The reported NDCG@20 are shown in Figure~\ref{fig:density_ml1m}. It is observed that DGCF consistently achieves the best performance, which indicates that DGCF can effectively propagate the information on bipartite graph. Moreover, we can find that the improvement that DGCF made on sparse user groups (i.e., $<35$ and $<65$) is higher (up to $24\%$) than that (up to $19\%$) for the dense groups (i.e., $<114$ and $<524$), demonstrating the DGCF can alleviate the cold-start issue by incorporating the CHP layers. Additionally, we observe that the performance of all models on the densest group are the best, which suggests that GCN-based models should perform better on a relative denser graph.

\subsection{Discussion on Drop-edge}
\begin{figure}
    \begin{subfigure}{0.24\textwidth}
    \includegraphics[width=.9\textwidth]{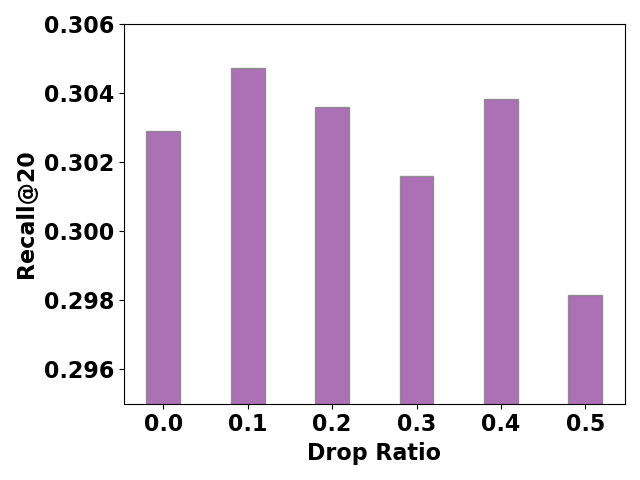}
    \caption{Recall@20}
    \label{fig:drop_recall}
    \end{subfigure}
    \hspace{-5mm}
    \begin{subfigure}{.24\textwidth}
    \includegraphics[width=.9\textwidth]{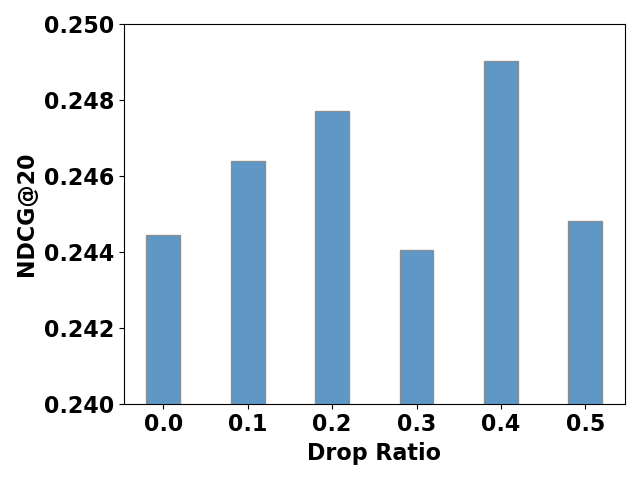}
    \caption{NDCG@20}
    \label{fig:drop_ndcg}
    \end{subfigure}
    \caption{Training with \textit{Drop-edge} on ML1M dataset.}
    \label{fig:drop edge}
\end{figure}
In this section, we conduct the experiments to study the effects of \textit{Drop-edge}~\cite{rong2019dropedge} operations. At each layer, we randomly drop a certain ratio of the edges from the original CHP laplacian matrix, i.e. the $(\mathcal{L} + \mathcal{L}_{c} + \mathbf{I})$. This operation is employed when training the parameters. During testing, we still use the original matrix to infer the user/item embeddings and make predictions. This drop-edge operation functions like a data augmentation method, which alleviates the oversmoothing problem and thus adds more robustness to the learned model. We conduct the experiment on ML1M dataset with a 4-layer DGCF model. We train the model with a certain ratio of drop-edges, and test its performance. The results are reported in Figure~\ref{fig:drop edge}. It is observed that dropping a certain ratio of the edges (i.e., prob. $> 0.0$), the testing performance becomes better. However, dropping too many edge leads to a worse performance. It breaks the original information propagation pattern, thus unable to well-fit the data. 

\subsection{Discussion on high-pass filtering}
\begin{figure}
    \begin{subfigure}{0.24\textwidth}
    \includegraphics[width=.9\textwidth]{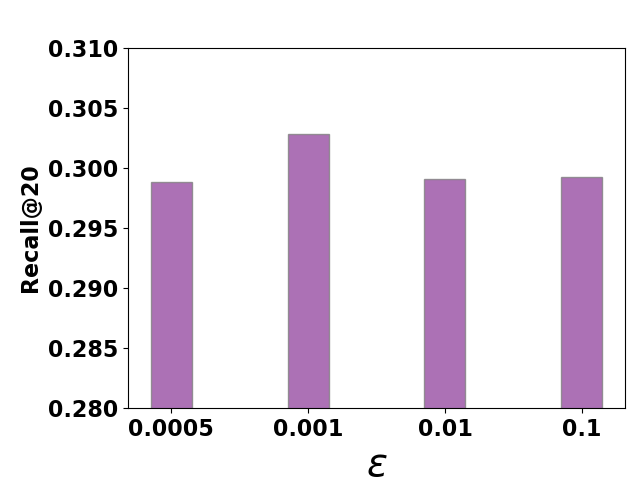}
    \caption{Recall@20}
    \label{fig:varepsilon_recall}
    \end{subfigure}
    \hspace{-5mm}
    \begin{subfigure}{.24\textwidth}
    \includegraphics[width=.9\textwidth]{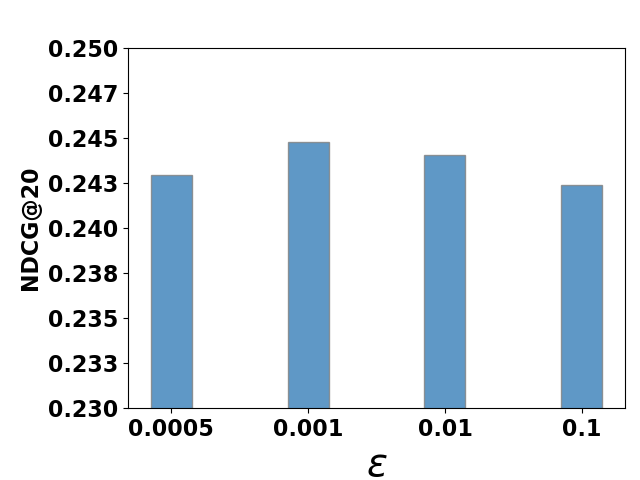}
    \caption{NDCG@20}
    \label{fig:varepsilon_ndcg}
    \end{subfigure}
    \caption{Performance w.r.t. $\varepsilon$ on ML1M dataset.}
    \label{fig:varepsilon_performance}
\end{figure}

In the Sec. 4.4. of the main paper, we introduce the high-pass filtering techniques. It uses a threshold $\varepsilon$ to filter cross-hop edges, which can not only increase the efficiency but also improve the performance. However, it is non-trivial to find an optimal $\varepsilon$. Smaller $\varepsilon$ filters fewer edges, which is vulnerable to oversmoothing, though solving the oscillation problem. Meanwhile, larger $\varepsilon$ filters more cross-hop edges, which spoils the ability of deoscillation and thus impairing the performance. To this end, we propose to set the $\varepsilon$ as a high-pass filtering threshold, such that the ratio\footnote{\label{note1}The ratio is defined as the greater one divided by the smaller one.} between the number of cross-hop edges and the number of direct bipartite graph edges is close to $1$. The number of cross-hop (CHP) and direct edges with respect to the value of $\varepsilon$ on ML1M dataset are presented in Table~\ref{tab:varepsilon_effect}. And, the corresponding performance with a $12$-layer DGCF model is reported in Figure~\ref{fig:varepsilon_performance}. We use a $12$-layer structure because it is easy to observe the oversmoothing problem. 

We set $\varepsilon=[10^{-1},10^{-2},10^{-3},5\times 10^{-4}]$. When $\varepsilon=10^{-1}$, there is no CHP edges, i.e., with only direct links from the original bipartite graph. Note that the number of direct edges is double of the user-item interactions as defined in Eq.~(5) of the main paper. When $\varepsilon$ decreases, more CHP edges are accepted for propagation. When $\varepsilon=10^{-3}$, the ratio is closest to $1$. In Figure~\ref{fig:varepsilon_performance}, it can be observed that its performance is the best compared with other values.  If keeping decreasing $\varepsilon$ to $5\times 10^{-4}$, too many cross edges are accepted. Therefore, the propagation consumes more time. And the model is vulnerable to oversmoothing. Hence, it performs worse compared with $\varepsilon=10^{-3}$.  

\begin{table}[!t]
\caption{Effects of $\varepsilon$ on ML1M dataset.}\label{tab:varepsilon_effect}
\begin{minipage}{1.0\columnwidth}
\begin{tabular}{lcccc}
\hline
 $\varepsilon$ =  &$10^{-1}$ &$10^{-2}$ &$10^{-3}$ &$5\times10^{-4}$ \\
 \hline
 \# direct edges & \multicolumn{4}{c}{296,598} \\
\hline

    \# cross edges &0 &1,752 &445,225  &2,614,123  \\
    ratio\footnotemark[2] & $\infty$ & 169.3 &1.6 & 8.8  \\ 
     time (s/epoch) &23.1 &23.2 &23.3 &60.3 \\ 
     \hline
\end{tabular}
\end{minipage}
\end{table}


%


\end{document}